\documentclass[reprint,amsmath,amssymb,aps,onecolumn]{revtex4-2}

\usepackage{graphicx}% Include figure files
\usepackage{bm}
\usepackage{hyperref}
\usepackage{graphicx}
\usepackage{braket}
\usepackage{amsfonts}
\usepackage{xcolor}
\begin{document}
\title{Breathers in solitonic room-temperature superlattice-induced superfluorescence
in quasi-2D perovskites}
\author{A. A. Gladkij}
\affiliation{Ioffe Physical-Technical Institute, Russian Academy of Sciences, St.
Petersburg 194021, Russia}
\author{N. N. Rosanov}
\affiliation{Ioffe Physical-Technical Institute, Russian Academy of Sciences, St.
Petersburg 194021, Russia}
\author{B. D. Fainberg}
\email{fainberg@hit.ac.il}
\affiliation{Faculty of Sciences, Physics Department, Holon Institute of Technology, 52
Golomb Street, POB 305, Holon 5810201, Israel}
\date{\today }

\begin{abstract}
Recently, a soliton mechanism for room-temperature superfluorescence in thin
perovskite films has been proposed, with a fundamental soliton predicted to
remain stable under LO phonon--exciton interactions. At the same time,
superlattice architectures offer a route to enhancing superfluorescence in
perovskites. Motivated by recent observations of room-temperature
superfluorescence in periodic superlattices of quasi-2D metal-halide
perovskites, we extend the 2D nonlocal nonlinear Schr\"{o}dinger equation
describing Wannier exciton--LO phonon interactions to superlattice structures,
obtaining a 3D nonlocal nonlinear Schr\"{o}dinger equation. We show that
interlayer tunnelling gives rise to breather dynamics corresponding to a
stable fundamental soliton in mixed coordinate--momentum space, with the
coordinate parallel to the layers and the momentum perpendicular to them. The
breather dynamics originate from miniband formation, which induces a
momentum-dependent phase modulation of the soliton. In the absence of
interlayer tunnelling, the breather dynamics disappear and the soliton becomes
stationary. These results establish a direct connection between interlayer
tunnelling, miniband formation and soliton dynamics, suggesting that breather
behavior can provide a signature of interlayer tunnelling in quasi-2D
perovskite superlattices.

\end{abstract}
\maketitle

\section*{INTRODUCTION}

Understanding the origin of stable macroscopic quantum coherence under
realistic room-temperature conditions is a fundamental challenge in modern
physics and is particularly relevant to the development of quantum
technologies, including quantum computing, communication and cryptography.
Recent experiments have demonstrated high-temperature superfluorescence (SF)
in hybrid perovskites
\cite{Gundogdu2022Nature_Phot,Sum_Mhaisalkar_Bruno2026AdvMat}, stimulating
considerable interest in the microscopic mechanisms that can sustain
macroscopic coherence in these materials. Significant progress has recently
been made through investigations of a soliton mechanism for high-temperature
SF, proposed independently in experimental \cite{Gundogdu2025Nature} and
theoretical \cite{s5h7-rpmk} studies (see also the commentary in
\cite{Zeng2026Materials_Futures}). The underlying system consists of a
quasi-two-dimensional (2D) Wannier exciton interacting with longitudinal
optical (LO) phonons through the Fr\"{o}hlich interaction, leading to the
formation of large polarons \cite{Fainberg_Osipov2024JCP,s5h7-rpmk}.
Low-energy, long-lived LO phonon modes couple to electronic transitions in
$CsPbBr_{3}$ \cite{Miyata17} and have lifetimes on the order of tens of
picoseconds. These modes may be associated with lead--halide--lead rocking
vibrations \cite{Gundogdu2025Nature}. Using the multiconfiguration Hartree
approach \cite{Osipov_Fainberg23PRB}, equations of motion for a single-exciton
wavefunction in momentum (wave-number $\mathbf{q}$) space were derived, in
which the LO phonon modes interact with the Wannier exciton through a
mean-field Hartree term \cite{Fainberg_Osipov2024JCP}. These nonlinear
equations were subsequently reduced to a 2D nonlocal nonlinear Schr\"{o}dinger
(NLS) equation \cite{s5h7-rpmk}, a framework widely used to describe nonlinear
wave propagation and soliton dynamics \cite{Krolikowski2000,Malomed2022}. This
formulation makes it possible to determine both the nonlinear solutions and
their stability, which is essential because 2D nonlocal NLS equations can
exhibit modulational instability (MI) \cite{Krolikowski2001,Krolikowsk2004}.
Two classes of stable solutions were identified: a plane-wave solution,
including the homogeneous distribution corresponding to the superradiant
state, and a fundamental soliton solution \cite{s5h7-rpmk}. The soliton regime
enhances the amplitude of the stable excitonic state, thereby promoting SF,
while reducing the spatial region over which the superradiant state is established.

Superfluorescence in perovskites can be further enhanced by incorporating
superlattice (SL) architectures
\cite{Raino2018Nature,Cherniukh2021Nature,Wildenborg2025ACSPhotonics}.
Recently, Wildenborg et al. observed room-temperature SF in the periodic SL
structures of quasi-2D metal-halide perovskites
\cite{Wildenborg2025ACSPhotonics}. Such structures consist of periodically
arranged, spatially separated emitting layers. The periodic arrangement
introduces coherent coupling between the layers and provides an additional
degree of freedom for controlling collective optical emission. In particular,
interlayer coupling can produce minibands and modify the spatial and phase
structure of collective excitonic states. In this context, it is also worth
noting the phonon-fluctuation synchronization effect, whereby delocalized
light--matter interactions in a multilayered material can shield
exciton--polaritons from phonon-induced decoherence
\cite{Mandal2026Nat_Commun}.

The formation of a SL raises a fundamental question concerning the fate of the
soliton mechanism established for an isolated quasi-2D layer. In particular,
it is necessary to determine how interlayer tunnelling modifies the stability
and dynamics of the fundamental soliton and whether new nonlinear states
emerge from the periodic structure. These questions are especially relevant
for interpreting room-temperature SF in quasi-2D perovskite superlattices,
where interlayer coupling may play an essential role in the collective dynamics.

Here, we generalize the 2D nonlocal NLS equation describing Wannier
exciton--LO phonon interactions to periodic superlattice structures, obtaining
a three-dimensional (3D) nonlocal NLS equation. We show that interlayer
tunnelling fundamentally modifies the soliton dynamics and gives rise to
breather states. These states correspond to the stable fundamental soliton in
mixed coordinate--momentum space, with the in-plane coordinate ($\mathbf{R}$)
describing motion parallel to the layers and the out-of-plane momentum
($K_{Z}$) describing motion perpendicular to them. The breather dynamics
originate from the formation of minibands, which produce a ($K_{Z}$)-dependent
phase modulation of the soliton. In the absence of interlayer tunnelling, the
miniband dispersion and associated phase modulation disappear, and the soliton
becomes stationary rather than exhibiting breather dynamics. Our results
therefore establish a direct connection between interlayer tunnelling,
miniband formation and nonlinear soliton dynamics, and suggest that the
observation of breather behavior could provide a signature of interlayer
tunnelling in quasi-2D perovskite superlattices.

\textbf{Exciton states in thin-film perovskite SLs and the model Hamiltonian.
}Consider a SL in direction $z$, which is perpendicular to $\mathbf{R}$.
Exciton state of a single thin film of thickness $l$ and area $L^{2}$ with the
lowest energy and the largest oscillator strength can be described by
\cite{Fainberg_Osipov2024JCP}%

\begin{equation}
\Psi_{\mathbf{q}11}(\mathbf{R},\mathbf{r},z_{e},z_{h})=\frac{1}{\sqrt{L^{2}}%
}\exp(i\mathbf{qR})\psi_{0}(\mathbf{r})f_{1}(z_{e})f_{1}(z_{h})
\label{eq:2Dwavefunction}%
\end{equation}
where $f_{m}(z_{i})=\sqrt{2/l}\sin(m\pi z_{i}/l)$; $i=e,h$; $\psi
_{0}(\mathbf{r})=\sqrt{8/(\pi a_{0}^{2})}\exp\left(  -2r/a_{0}\right)  $
refers to the normalized hydrogen atom wave function, $a_{0}$ represents the
exciton Bohr radius. Here $\mathbf{R}$ and $\mathbf{r}$ represent the exciton
coordinates in the film plane, describing the center-of-mass motion and the
electron-hole relative motion, respectively.

In a SL with a period $d$, such as quasi-two-dimensional perovskite
structures, the Wannier exciton wavefunction exhibits a modulated
Bloch-periodic structure arising from the periodic potential along the growth
direction (typically the $z$-axis). Due to the periodicity along the
$z$-direction, the center-of-mass motion obeys Bloch's theorem. In thin-film
perovskites (with free in-plane motion and quantum confinement along the
$z$-direction), one can write the basis function $\Psi_{\mathbf{q}}%
(\mathbf{R},\mathbf{r},z_{e},z_{h})$ as%
\begin{equation}
\Psi_{\mathbf{q}}(\mathbf{R},\mathbf{r},z_{e},z_{h})=\frac{1}{\sqrt{L^{2}}%
}\exp(i\mathbf{qR})\psi_{0}(\mathbf{r})\chi_{e}(z_{e})\chi_{h}(z_{h})
\label{eq:Psi_q2}%
\end{equation}
Integrating both sides of the last equation with respect to the relative
coordinate $z=z_{e}-z_{h}$, we arrive at a new basis function%
\begin{equation}
\Psi_{\mathbf{q}}(\mathbf{R},\mathbf{r},Z)=\frac{1}{\sqrt{L^{2}}}%
\exp(i\mathbf{qR})\psi_{0}(\mathbf{r})|K_{Z}\rangle=|\mathbf{q}\rangle
|K_{Z}\rangle\label{eq:Psi_q}%
\end{equation}
where $|\mathbf{q}\rangle=\frac{1}{\sqrt{L^{2}}}\exp(i\mathbf{qR})\psi
_{0}(\mathbf{r})$ \cite{Fainberg_Osipov2024JCP},
\begin{equation}
\left\vert K_{z}\right\rangle =%
%TCIMACRO{\dsum \limits_{n}}%
%BeginExpansion
{\displaystyle\sum\limits_{n}}
%EndExpansion
e^{iK_{z}nd}\int dz\chi_{e}(Z+p_{e}z)\chi_{h}(Z-p_{h}z) \label{eq:ksi(Z)}%
\end{equation}
$Z=(m_{e}^{\ast}z_{e}+m_{h}^{\ast}z_{h})/(m_{e}^{\ast}+m_{h}^{\ast})$ is the
coordinate of the centre of mass of an electron and a hole in direction $z$,
$m_{e}^{\ast}$ and $m_{h}^{\ast}$ is the electron and hole effective mass,
respectively; $K_{Z}$ are wave vectors in the $Z$-direction, $p_{i}\equiv
m_{r}/m_{i}^{\ast}$, $1/m_{r}=1/m_{e}^{\ast}+1/m_{h}^{\ast}$ is the inverse
reduced mass.

In deriving Eq.(\ref{eq:Psi_q}), we adopted the adiabatic approximation.
Specifically, when the superlattice period $d$ is much larger than the exciton
radius ($a_{0}\ll d$), the SL potential varies slowly on the exciton length
scale, allowing the relative motion to be treated as parametrically dependent
on the center-of-mass coordinate $Z$.

The Hamiltonian for a Wannier exciton interacting with LO phonons consists of
three components: the Wannier exciton energy operator ($\hat{H}_{exc}^{W}$),
the phonon energy operator ($\hat{H}_{ph}$), and the Wannier exciton-LO phonon
interaction energy operator ($\hat{H}_{I}^{W}$) \cite{Fainberg_Osipov2024JCP}
\begin{equation}
\hat{H}_{0}=\hat{H}_{ex}^{W}+\hat{H}_{ph}+\hat{H}_{I}^{W} \label{eq.:H_03}%
\end{equation}
where
\begin{equation}
\hat{H}_{exc}^{W}=\hbar\sum_{\mathbf{k}}W(\mathbf{k})B_{\mathbf{k}}^{\dag
}B_{\mathbf{k}}, \label{eq:Wex_hamilt4}%
\end{equation}%
\begin{equation}
\hat{H}_{ph}=\sum_{\mathbf{\mathbf{q}}}\hbar\omega_{l}b_{\mathbf{q}}^{\dag
}b_{\mathbf{q}}, \label{eq.:H_ph}%
\end{equation}

\begin{equation}
\hat{H}_{I}^{W}=-i\hbar\sum_{\mathbf{k\mathbf{q}}}D_{l}^{pol}(\mathbf{k}%
;00)B_{\mathbf{k+q}}^{\dag}B_{\mathbf{q}}(b_{\mathbf{k}}-b_{\mathbf{-k}}%
^{\dag}), \label{eq.:H^pol(W)_Ia}%
\end{equation}
$B_{\mathbf{q}}^{\dag}$($B_{\mathbf{q}}$) and $b_{\mathbf{k}}^{\dag}%
$($b_{\mathbf{k}}$) are creation (annihilation) operators of excitons and
phonons, respectively. Here $W(\mathbf{k})=E_{c}(\mathbf{k}_{0})-E_{v}%
(\mathbf{k}_{0})+E_{0}+\frac{\hbar^{2}k^{2}}{2M^{\ast}}$, $E_{0}$ is the
energy of the internal structure of an exciton in state $n=0$; $E_{c}%
(\mathbf{k}_{0})-E_{v}(\mathbf{k}_{0})$ is the energy gap between the minimum
of the first conduction band and the maximum of the valence band; $M^{\ast
}=m_{e}^{\ast}+m_{h}^{\ast}$, where $m_{e}^{\ast}$ and $m_{h}^{\ast}$ are the
electron and hole effective masses, respectively; $D_{l}^{pol}(\mathbf{k};00)$
is the electronic matrix element for coupling function of the Wannier
exciton-LO phonon interaction \cite{Fainberg_Osipov2024JCP}.

It is worth noting that, in the case of SL, $B_{\mathbf{q}}=|0\rangle
\langle\mathbf{q}|\langle K_{Z}|$, while the Wannier exciton-LO phonon
interaction energy operator $\hat{H}_{I}^{W}$ retains the form given by
Eq.(\ref{eq.:H^pol(W)_Ia}). Due the presence of SL, the exciton spectrum forms
minibands, and the exciton energy with wavevector $\mathbf{k}$ in the thin
film plane in a hydrogen-atom-like internal electronic energy state $n=0$ is
given by:
\begin{equation}
E_{exc}(\mathbf{k},K_{Z})=\hbar W(\mathbf{k})+E_{SL}\left(  K_{Z}\right)
\label{eq:W_n}%
\end{equation}
Then the Wannier exciton energy operator in a SL takes a form:%
\begin{equation}
\hat{H}_{exc}^{W}=\sum_{K_{Z}\mathbf{k}}E_{exc}(\mathbf{k},K_{Z}%
)B_{\mathbf{k}}^{\dag}B_{\mathbf{k}} \label{eq.:H^W_exc}%
\end{equation}

The basis set of single-exciton states is $|\mathbf{q},K_{Z}\rangle\equiv
\Psi_{\mathbf{q}}(\mathbf{R},\mathbf{r},K_{Z})=B_{\mathbf{q}}^{\dag}|0\rangle$
with energies $E_{exc}(\mathbf{q},K_{Z})$. A wave function of the system can
be written as
\begin{equation}
|\Psi(t)\rangle=\sum_{\mathbf{q}K_{Z}}[G(\mathbf{q},K_{Z}|t)|0\rangle
+C_{0}(\mathbf{q},K_{Z}|t)|K_{Z}\rangle|\mathbf{q}\rangle] \label{eq:Psi(t)}%
\end{equation}
where $|\Psi(t)\rangle$ is normalized. This gives
\begin{equation}
\sum_{\mathbf{q}K_{Z}}[|C_{0}(\mathbf{q},K_{Z}|t)|^{2}+|G(\mathbf{q}%
,K_{Z}|t)|^{2}]=1 \label{eq:norm}%
\end{equation}

It is convenient to formulate the equations in terms of the complex-valued
polarization $\langle B_{\mathbf{q}}\rangle=\langle\Psi(t)|B_{\mathbf{q}}%
|\Psi(t)\rangle$ where%
\begin{equation}
\langle B_{\mathbf{q}}\rangle=C_{0}(\mathbf{q},K_{Z}|t)\sum_{\mathbf{q}%
^{\prime}K_{Z}^{\prime\prime}}G^{\ast}(\mathbf{q}^{\prime},K_{Z}^{\prime
\prime}|t) \label{eq.:<B_q>}%
\end{equation}

In the first approximation, when the population of the ground state is close
to $1$, one can set $\sum_{\mathbf{q}^{\prime}K_{Z}^{\prime\prime}}G^{\ast
}(\mathbf{q}^{\prime},K_{Z}^{\prime\prime}|t)\approx1$, and $\langle
B_{\mathbf{q}}\rangle\simeq C_{0}(\mathbf{q},K_{Z}|t)$. In other words, the
complex-valued polarization is close to the coefficient $C_{0}(\mathbf{q}%
,K_{Z}|t)$.

\section*{RESULTS}

\subsection*{Nonlocal nonlinear Schr\"odinger equation for a SL}

We first extend the theoretical description of the exciton--LO phonon
interaction developed for quasi-2D perovskite layers to a periodic SL. The
Heisenberg equations for the phonon $b_{\mathbf{q}}$ and exciton
$B_{\mathbf{q}}$ operators are averaged using the single-exciton wavefunction
and the corresponding normalization condition, Eq.(\ref{eq:norm}), in the
limit in which the ground-state population remains close to unity when the
excitonic polarization $\langle B_{\mathbf{q}}\rangle$ can be approximated by
the coefficient $C_{0}(\mathbf{q},K_{Z}|t)$ of the single-exciton
wavefunction. As a result, we get a nonlinear equation,
Eq.(\ref{eq.:dC_0(q,K_Z,t)/dt}) in Section \textbf{METHODS}, for the
coefficients $C_{0}(\mathbf{q},K_{Z}|t)$ in the momentum space.

Applying a Fourier transform with respect to the in-plane momentum
$\mathbf{q}$ to Eq.(\ref{eq.:dC_0(q,K_Z,t)/dt}) for $C_{0}(\mathbf{q}%
,K_{Z}|t)\equiv C(\mathbf{q},K_{Z}|t)$, using
\[
C(\mathbf{q},K_{Z}|t)=\frac{1}{\sqrt{N}}%
%TCIMACRO{\dsum \limits_{m}}%
%BeginExpansion
{\displaystyle\sum\limits_{m}}
%EndExpansion
C_{m}(K_{Z},t)\exp(-i\mathbf{qR}_{m}),C_{m}(K_{Z},t)=\frac{1}{\sqrt{N}}%
%TCIMACRO{\dsum \limits_{\mathbf{q}}}%
%BeginExpansion
{\displaystyle\sum\limits_{\mathbf{q}}}
%EndExpansion
C(\mathbf{q},K_{Z}|t)\exp(i\mathbf{qR}_{m}),
\]
where $N$ is the number of particles (electron-hole pairs) in a thin layer
\cite{Fainberg_Osipov2024JCP}, gives an equation for the exciton amplitude
$C_{m}(K_{Z},t)$ in the plane of the layers:
\begin{align}
i\frac{d}{dt}C_{m}(K_{Z},t)  &  =\left[  \bar{W}_{0}+\frac{1}{\hbar}%
E_{SL}\left(  K_{Z}\right)  -\frac{\hbar}{2M^{\ast}}\frac{\partial^{2}%
}{\partial\mathbf{R}_{m}^{2}}\right]  C_{m}(K_{Z},t)\nonumber\\
&  -2\sum_{K_{Z^{\prime}}}%
%TCIMACRO{\dsum \limits_{m^{\prime}}}%
%BeginExpansion
{\displaystyle\sum\limits_{m^{\prime}}}
%EndExpansion
|C_{m^{\prime}}(K_{Z^{\prime}})|^{2}\omega_{0}(R_{m^{\prime}m})C_{m}(K_{Z},t)
\label{eq.:dC_m(x,y,nd,t)/dt_super}%
\end{align}
where $\omega_{0}(R_{m^{\prime}m})$ is the Fourier transform of $\omega
_{0}(k)=\frac{\omega_{l}}{\omega_{l}^{2}+\gamma^{2}}|D_{l}^{pol}%
(\mathbf{k};00)|^{2}$ \cite{s5h7-rpmk}. Eq.(\ref{eq.:dC_m(x,y,nd,t)/dt_super})
has the form of a nonlocal nonlinear Schr\"{o}dinger (NLS) equation, with the
superlattice dispersion entering through the term $E_{SL}(K_{Z})$.

Eq.(\ref{eq.:dC_m(x,y,nd,t)/dt_super}) can be written for the amplitudes
$\tilde{C}_{m}(K_{Z},t)=C_{m}(K_{Z},t)\exp[i(\bar{W}_{0}+\frac{1}{\hbar}%
E_{SL}\left(  K_{Z}\right)  )t]$ in the polar coordinates and the continuum
approximation for a radially symmetric in-plane distribution as follows:%
\begin{align}
\{  &  i\frac{\partial}{\partial t}+\frac{\hbar}{2M^{\ast}}\left[  \frac{1}%
{R}\frac{\partial}{\partial R}\left(  R\frac{\partial}{\partial R}\right)
+\frac{1}{R^{2}}\frac{\partial^{2}}{\partial\varphi^{2}}\right]
+\sum_{K_{Z^{\prime}}}\frac{4\pi lG_{0}}{l_{0}^{3}}\exp\left(  -\frac{R^{2}%
}{\rho^{2}}\right)  \int_{0}^{\infty}dR^{\prime}R^{\prime}\left\vert \tilde
{C}(K_{Z^{\prime}},R^{\prime},\varphi,t)\right\vert ^{2}\nonumber\\
&  \times\exp\left(  -\frac{R^{\prime2}}{\rho^{2}}\right)  \left[  \left(
1-\frac{R^{\prime2}+R^{2}}{\rho^{2}}\right)  I_{0}\left(  \frac{2R^{\prime}%
R}{\rho^{2}}\right)  +\frac{2R^{\prime}R}{\rho^{2}}I_{1}\left(  \frac
{2R^{\prime}R}{\rho^{2}}\right)  \right]  \}\tilde{C}(K_{Z},R,\varphi
,t)\nonumber\\
&  =0 \label{eq.:dC_(K_Z,R,fi,t)/dt}%
\end{align}
where $l_{0}^{3}$ is the volume of the unit cell, $I_{n}(A)$ is the modified
Bessel functions of the first kind. In deriving
Eq.(\ref{eq.:dC_(K_Z,R,fi,t)/dt}), we used the approximation for the function
$\omega_{0}(R)$ from Ref.\cite{s5h7-rpmk}, and $G_{0}=\omega_{0}(R=0)$.
Eq.(\ref{eq.:dC_(K_Z,R,fi,t)/dt}) is the SL generalization of the 2D nonlocal
NLS equation previously derived for an isolated quasi-2D perovskite layer
\cite{s5h7-rpmk}. Because the nonlinear term retains the same nonlocal form as
in the isolated-layer problem, the superlattice equation supports the same
class of stable fundamental-soliton solutions in the in-plane coordinate.
Fig.~\ref{fig:C(K_Z,R,t)} shows the resulting radially symmetric profiles of
$\tilde{C}(K_{Z},R,t)$ for $\tilde{N}=15$. The profiles corresponding to
different $K_{Z}$ values have identical radial shapes, and their superposition
reproduces the same radial soliton profile. The same behavior is obtained for
$\tilde{N}=25$ layers (Fig.~\ref{fig:C(K_Z,R,t)25}). This result follows from
the fact that the nonlinear equation for $\tilde{C}(K_{Z},R,t)$ is symmetric
with respect to $K_{Z}$.

The soliton, however, acquires a qualitatively different character along the
direction perpendicular to the layers. In an isolated quasi-2D layer, the
fundamental soliton is localized in real space. In the SL, it is more
naturally described in mixed coordinate--momentum space: it remains localized
in the in-plane coordinate $R$, whereas the perpendicular degree of freedom is
represented by the momentum $K_{Z}$. The relation between the original
amplitude and $\tilde{C}$ contains the superlattice dispersion,
\begin{equation}
C(K_{Z},R,\varphi,t)=\tilde{C}(K_{Z},R,\varphi,t)\exp[-i(\bar{W}_{0}%
+E_{SL}\left(  K_{Z}\right)  )t] \label{eq.:C_(K_Z,R,fi,t)}%
\end{equation}
and therefore introduces a $K_{Z}$-dependent phase when interlayer tunnelling
is present (see below). This phase is the origin of the subsequent dynamics in
real space.

\subsection*{Miniband formation and interlayer tunnelling}

\begin{figure}
%[ptbptbptb]
\begin{center}
\includegraphics[scale=0.315]{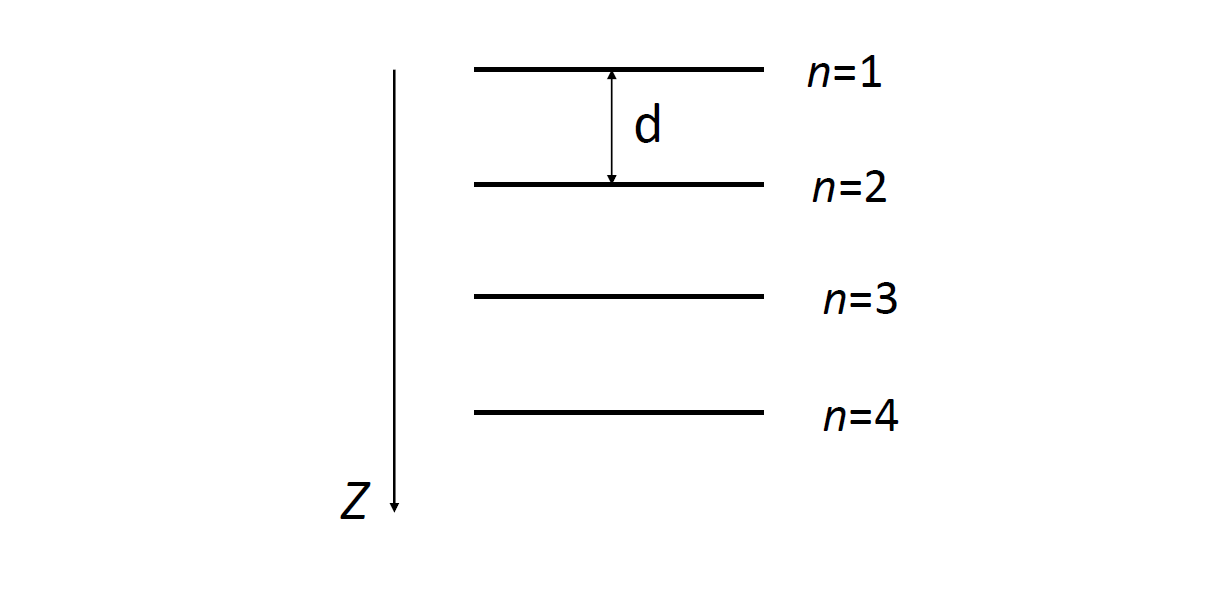}%
\caption{Frenkel exciton model of SL with period $d$.}%
\label{fig:Frenkel}%
\end{center}
\end{figure}

To describe the motion perpendicular to the layers in SL, we use a Frenkel
exciton model. Each quasi-2D exciton localized in a layer is treated as an
effective molecular excitation, while tunnelling between neighboring layers is
characterized by the coupling parameter $J\simeq J_{e}+J_{h}$. Restricting the
interaction to nearest-neighbour layers gives the standard one-dimensional
aggregate Hamiltonian \cite{Fra98,Che63,Spano91,Muk95,Fainberg2016CP}. Its
eigenstates are labelled by the discrete wave vector $K_{Z}=1,...,\tilde{N}$:
\begin{equation}
\left\vert K_{Z}\right\rangle =\sqrt{\frac{2}{\tilde{N}+1}}\sum_{n=1}%
^{\tilde{N}}\sin\left(  \frac{\pi nK_{Z}}{\tilde{N}+1}\right)  \left\vert
ne\right\rangle \label{eq:|K_Z>}%
\end{equation}
where $\tilde{N}$ is the number of layers in SL, and have the dispersion%

\begin{equation}
E_{SL}\left(  K_{Z}\right)  =2\hbar J\cos\left(  \frac{\pi K_{Z}}{\tilde{N}%
+1}\right)  . \label{eq:E_K_Z}%
\end{equation}
Thus, interlayer tunnelling produces a miniband whose width is controlled by
$J$. The eigenstates are delocalized over the layers, whereas the
corresponding layer states $\left\vert ne\right\rangle $ are obtained through
a sine Fourier transform of the $K_{Z}$ representation: $\left\vert
ne\right\rangle =\sqrt{\frac{2}{\tilde{N}+1}}\sum_{K_{Z}=1}^{N}\sin\left(
\frac{\pi nK_{Z}}{\tilde{N}+1}\right)  \left\vert K_{Z}\right\rangle $, and
$n$-th exciton is located at the coordinate $Z=nd$ where $d$ is a period of
the superlattice (see Fig.\ref{fig:Frenkel}).

In the coordinate space, we will work in the basis of the exciton wave
functions $\left\vert \mathbf{R},K_{Z}\right\rangle =\left\vert \mathbf{R}%
\right\rangle \left\vert K_{Z}\right\rangle $. Then the wave function of the
system under consideration can be written as $|\Psi_{e}(t)\rangle
=\sum_{\mathbf{R}K_{Z}}C(K_{Z},\mathbf{R},t)|\mathbf{R}\rangle|K_{Z}\rangle$.
To calculate the soliton amplitude in layer $n$, $C(nd,R,\varphi,t)$, we will
consider a quantity $\left\langle \mathbf{R},ne\right\vert \Psi_{e}%
(t)\rangle=\left\langle \mathbf{R}|\langle ne\right\vert \sum_{\mathbf{R}%
^{\prime}K_{Z}}C(\mathbf{R}^{\prime},K_{Z},t)|\mathbf{R}^{\prime}\rangle
|K_{Z}\rangle$. Using Eq.(\ref{eq:|K_Z>}), we get%
\begin{equation}
\langle\mathbf{R},ne|\Psi_{e}(t)\rangle=\sqrt{\frac{2}{\tilde{N}+1}}%
\sum_{K_{Z}}C(\mathbf{R},K_{Z},t)\sin\left(  \frac{\pi nK_{Z}}{\tilde{N}%
+1}\right)  \label{eq:C(nd,R,t)}%
\end{equation}
where $\sqrt{\frac{2}{\tilde{N}+1}}\sum_{K_{Z}}C(\mathbf{R},K_{Z}%
,t)\sin\left(  \frac{\pi nK_{Z}}{\tilde{N}+1}\right)  $ can be considered as
amplitudes $C(nd,R,\varphi,t)$. In other words, $C(nd,R,\varphi,t)$ is a sine
Fourier-transform of $C(\mathbf{R},K_{Z},t)$. This relation provides a direct
connection between the $K_{Z}$-dependent phase acquired from the miniband
dispersion and the time-dependent distribution of the soliton among the layers.

\subsection*{Breather dynamics induced by interlayer tunnelling}

\begin{figure}
%[ptb]
\begin{center}
\includegraphics[scale=0.435]{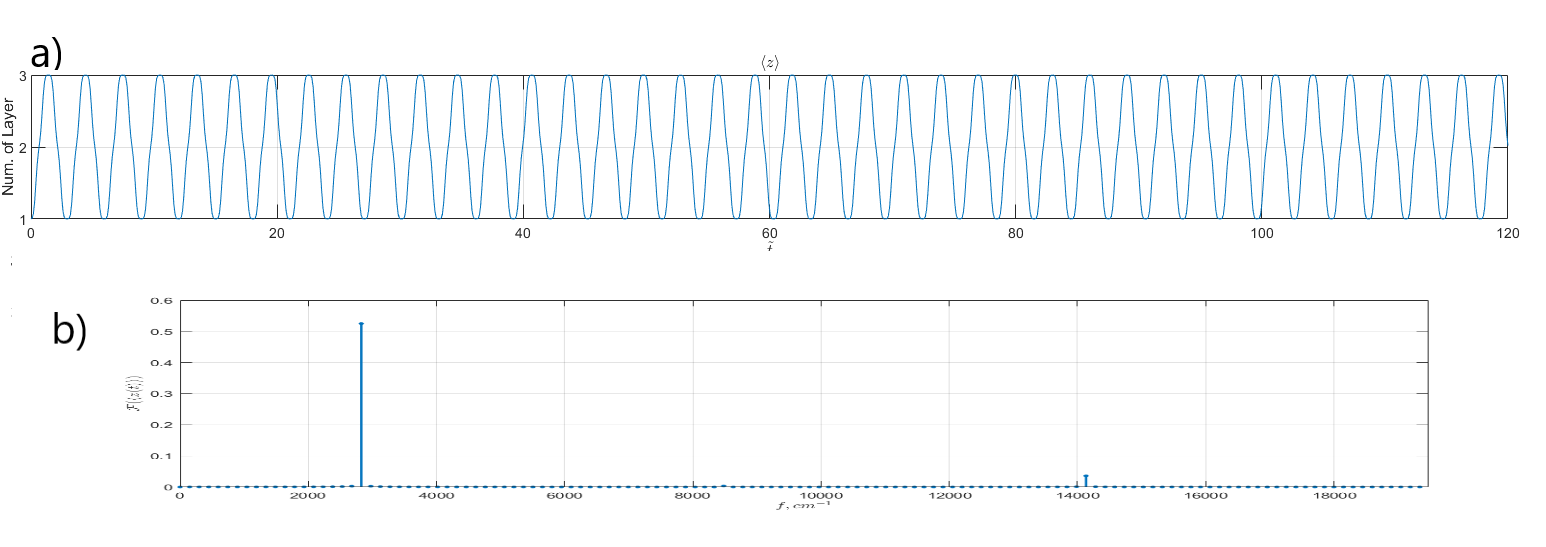}%
\caption{Average coordinate $\langle Z(t)\rangle$ as a function of time (a)
and the spectrum of $\langle Z(t)\rangle$ (b) for $J=2000$ $cm^{\mathbf{-}1}%
$and $\tilde{N}=3$.}%
\label{fig:Breather}%
\end{center}
\end{figure}

We next examine the consequences of the miniband dispersion for the real-space
dynamics. For $\tilde{N}=3$ and $J=2000$~cm$^{-1}$, the maximum value of
$C(nd,R,\varphi,t)$ exhibits pronounced temporal oscillations, particularly in
the outer layers (Fig.~\ref{fig:C(nd,R,t)}). These oscillations indicate a
periodic redistribution of the soliton amplitude among the layers that is
characteristic for a breather. To characterize this motion independently of
the local layer amplitudes, we calculate the average position $\langle
Z(t)\rangle$. For $\tilde{N}=3$, $\langle Z(t)\rangle$ exhibits clear periodic
oscillations, with a dominant frequency of approximately $860$~cm$^{-1}$
(Fig.~\ref{fig:Breather}).

\begin{figure}
%[ptbptb]
\begin{center}
\includegraphics[scale=0.39]{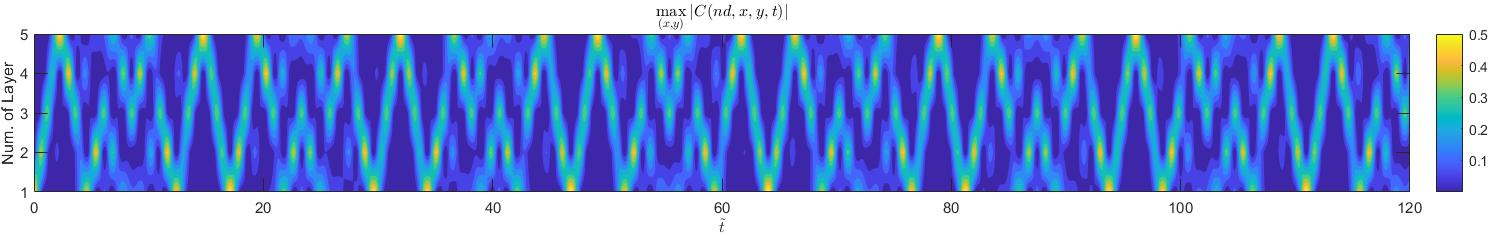}%
\caption{A maximum value of $C(nd,R,\varphi,t)$ as a function of time for
$J=2000$ $cm^{\mathbf{-}1}$and $\tilde{N}=5$.}%
\label{fig:5layers}%
\end{center}
\end{figure}

\begin{figure}
%[ptbptbptb]
\begin{center}
\includegraphics[scale=0.39]{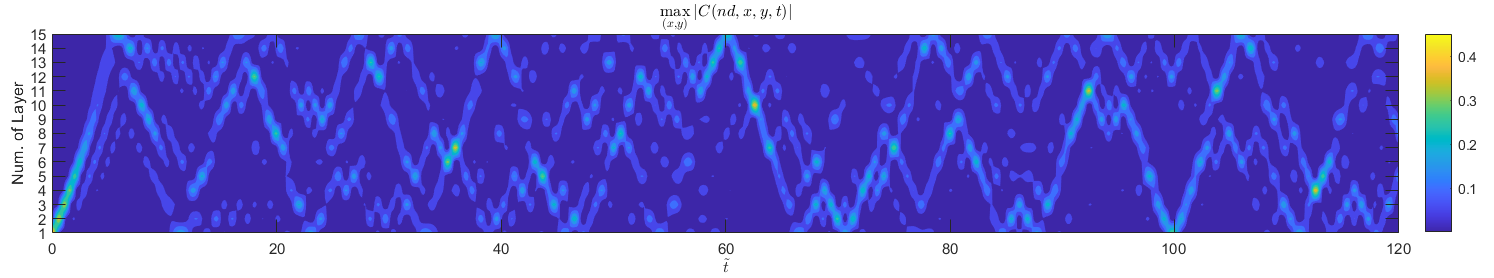}%
\caption{A maximum value of $C(nd,R,\varphi,t)$ as a function of time for
$J=2000$ $cm^{\mathbf{-}1}$and $\tilde{N}=15$.}%
\label{fig:15layers}%
\end{center}
\end{figure}

\begin{figure}
%[ptbptbptbptb]
\begin{center}
\includegraphics[scale=0.402]{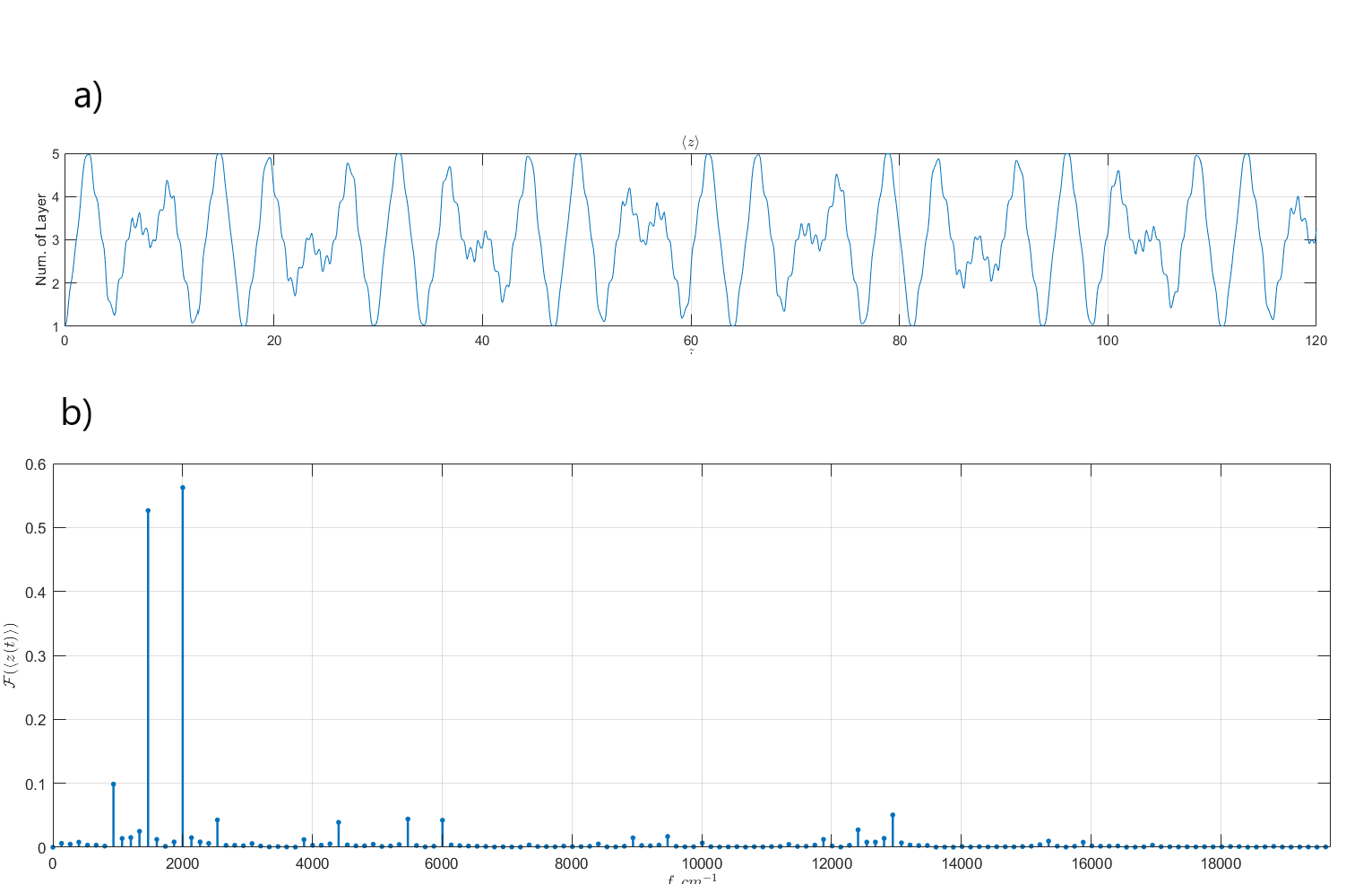}%
\caption{Average coordinate $\langle Z(t)\rangle$ as a function of time (a)
and the spectrum of $\langle Z(t)\rangle$ (b) for $J=2000$ $cm^{\mathbf{-}1}%
$and $\tilde{N}=5$.}%
\label{fig:Breather5}%
\end{center}
\end{figure}

\begin{figure}
%[ptbptbptbptbptb]
\begin{center}
\includegraphics[scale=0.351]{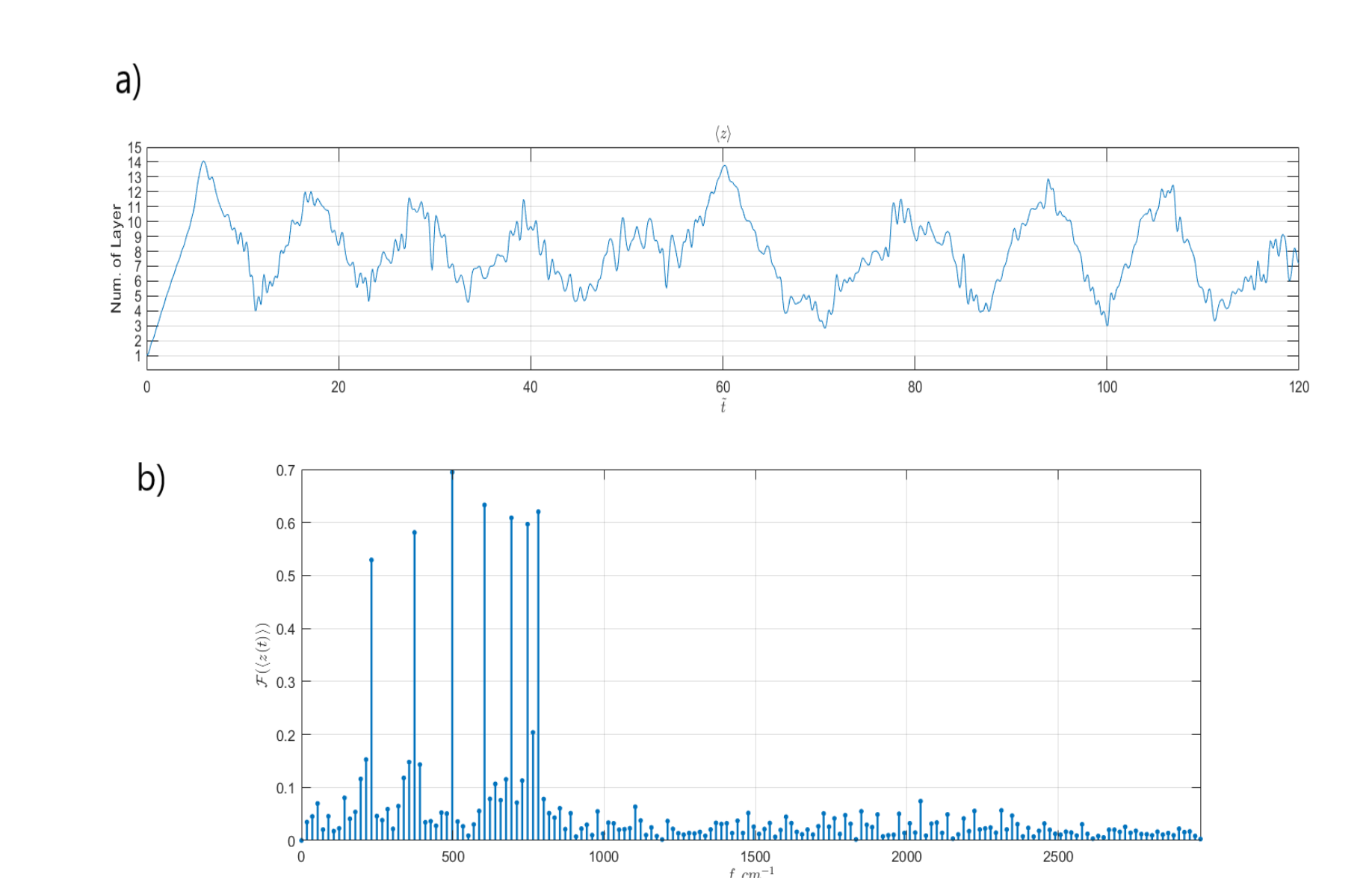}%
\caption{Average coordinate $\langle Z(t)\rangle$ as a function of time (a)
and the spectrum of $\langle Z(t)\rangle$ (b) for $J=2000$ $cm^{\mathbf{-}1}%
$and $\tilde{N}=15$.}%
\label{fig:Breather15}%
\end{center}
\end{figure}

The temporal dynamics become richer as the number of layers increases. For
$\tilde{N}=5$ and $\tilde{N}=15$, the layer-resolved amplitudes display
increasingly complex oscillations, with substantial amplitudes appearing in
intermediate layers in addition to the outer layers (Figs.~\ref{fig:5layers}
and \ref{fig:15layers}). The corresponding $\langle Z(t)\rangle$ remains
oscillatory, but its spectrum contains several frequencies
(Figs.~\ref{fig:Breather5} and \ref{fig:Breather15}). The resulting dynamics
can be understood as beating between different oscillatory contributions.
Thus, increasing the number of layers does not destroy the coherent
oscillatory motion; rather, it enriches its frequency content.

The origin of these oscillations can be seen directly from the SL dispersion.
Transforming the miniband energy from the $K_{Z}$ representation to the layer
coordinate $Z=nd$ and treating $Z$ as a continuous variable produces an
effective $Z$-dependent energy landscape:
\begin{equation}
\left\langle ne\right\vert E_{SL}\left\vert ne\right\rangle =\frac{\hbar
J}{2(\tilde{N}+1)}\left[  2-\frac{\sin[\frac{\pi}{2}\frac{(2\tilde{N}%
+1)}{\tilde{N}+1}(2n+1)]}{\sin[\frac{\pi}{2}\frac{(2n+1)}{\tilde{N}+1}]}%
-\frac{\sin[\frac{\pi}{2}\frac{(2\tilde{N}+1)}{\tilde{N}+1}(2n-1)]}{\sin
[\frac{\pi}{2}\frac{(2n-1)}{\tilde{N}+1}]}\right]  \label{eq:E_SL(n)}%
\end{equation}

\begin{figure}
%[ptbptbptbptbptbptb]
\begin{center}
\includegraphics[scale=1.008]{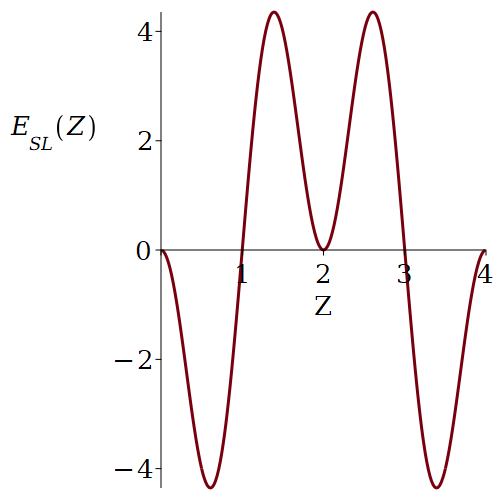}%
\caption{Energy dependence on coordinate $Z$ for $\tilde{N}=3$.}%
\label{fig:E_n3}%
\end{center}
\end{figure}

\begin{figure}
%[ptbptbptbptbptbptbptb]
\begin{center}
\includegraphics[scale=1.008]{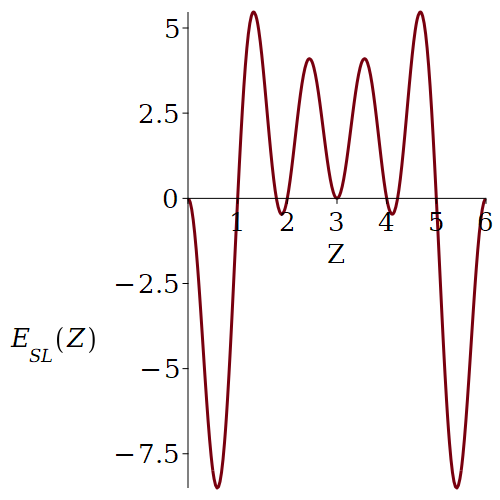}%
\caption{Energy dependence on coordinate $Z$ for $\tilde{N}=5$.}%
\label{fig:E_n5}%
\end{center}
\end{figure}

\begin{figure}
%[ptbptbptbptbptbptbptbptb]
\begin{center}
\includegraphics[scale=1.008]{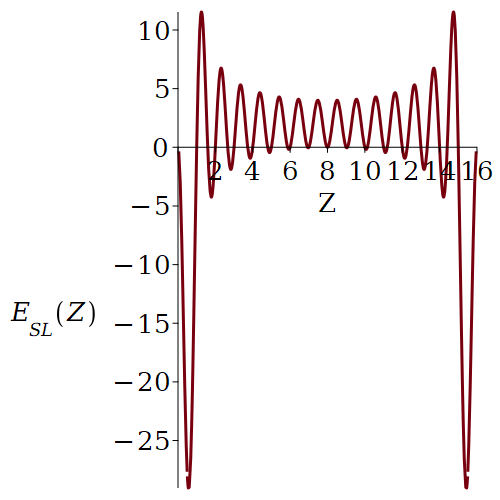}%
\caption{Energy dependence on coordinate $Z$ for $\tilde{N}=15$.}%
\label{fig:E_n15}%
\end{center}
\end{figure}

As the number of layers $\tilde{N}$ increases, this landscape becomes more
structured (Figs.~\ref{fig:E_n3}, \ref{fig:E_n5} and \ref{fig:E_n15}), leading
to increasingly complex motion of the soliton in the layer direction. The
resulting oscillations are therefore not an additional nonlinear instability
of the fundamental soliton. Rather, they arise from the coherent phase
evolution generated by the interlayer tunnelling and the associated miniband dispersion.

\subsection*{Dependence of the breather dynamics on the tunnelling strength}

\begin{figure}
%[ptbptbptbptbptbptbptbptbptb]
\begin{center}
\includegraphics[scale=0.552]{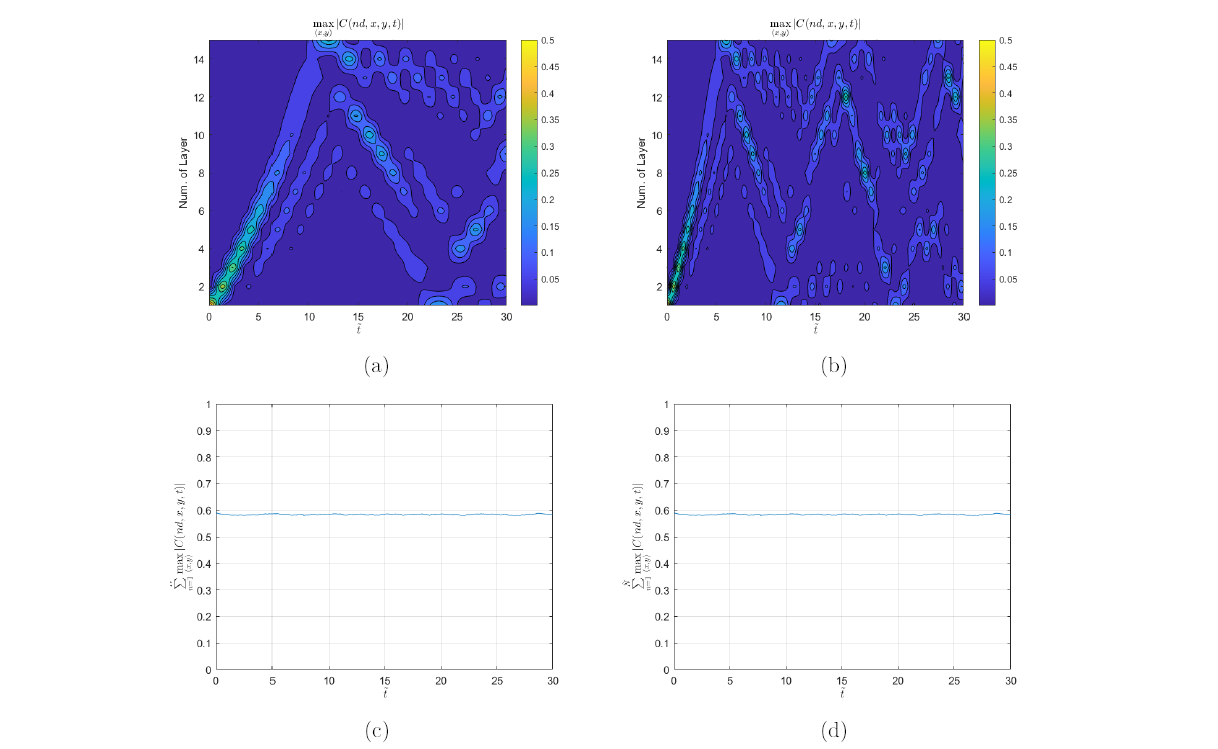}%
\caption{A maximum value of $C(nd,R,\varphi,t)$ as a function of time for
$J=1000$ $cm^{\mathbf{-}1}$ (a) and $J=2000$ $cm^{\mathbf{-}1}$ (b) when
$\tilde{N}=15$. Dependence of the sum of the maximum values of $C(nd,R,\varphi
,t)$ over the layers as a function of time for $J=1000$ $cm^{\mathbf{-}1}$ (c)
and $J=2000$ $cm^{\mathbf{-}1}$ (d).}%
\label{fig:J15}%
\end{center}
\end{figure}

\begin{figure}
%[ptbptbptbptbptbptbptbptbptbptb]
\begin{center}
\includegraphics[scale=0.6227]{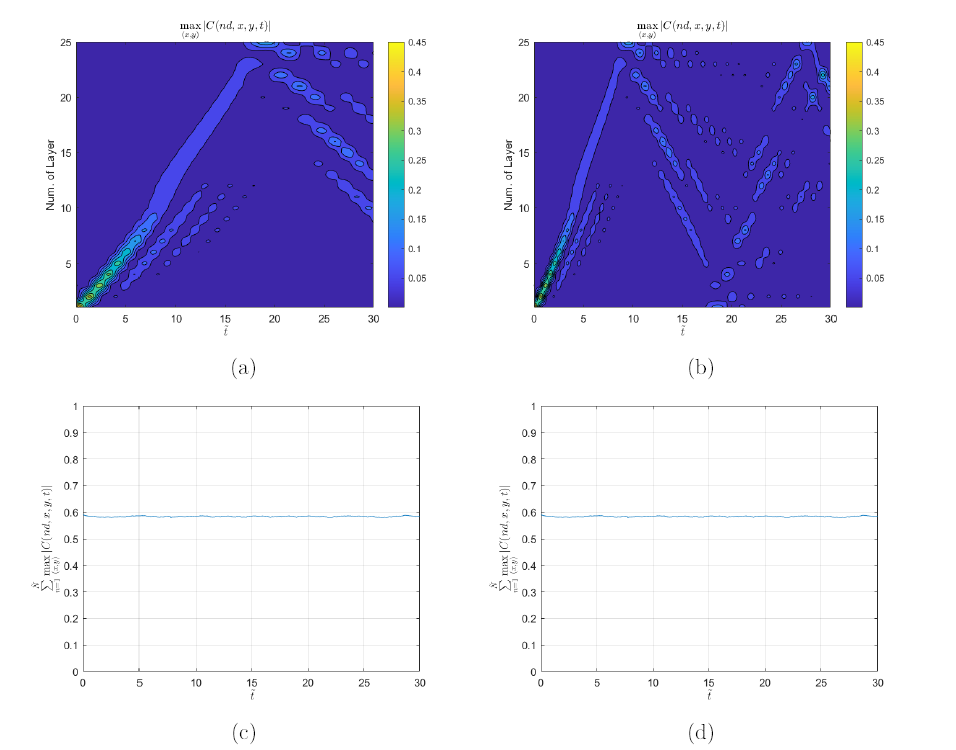}%
\caption{A maximum value of $C(nd,R,\varphi,t)$ as a function of time for
$J=1000$ $cm^{\mathbf{-}1}$ (a) and $J=2000$ $cm^{\mathbf{-}1}$ (b) when
$\tilde{N}=25$. Dependence of the sum of the maximum values of $C(nd,R,\varphi
,t)$ over the layers as a function of time for $J=1000$ $cm^{\mathbf{-}1}$ (c)
and $J=2000$ $cm^{\mathbf{-}1}$ (d).}%
\label{fig:J25}%
\end{center}
\end{figure}

We next investigate how the breather dynamics depend on the interlayer
tunnelling strength. Figures~\ref{fig:J15} and \ref{fig:J25} show the maximum
value of $C(nd,R,\varphi,t)$ as a function of time for $\tilde{N}=15$ and
$25$, respectively, for $J=1000$ and $2000$~cm$^{-1}$. The temporal modulation
of the layer-resolved amplitudes demonstrates that the characteristic dynamics
persist over a broad range of tunnelling strengths. Increasing $J$ changes the
phase accumulated from the miniband dispersion and therefore modifies the
temporal evolution of the layer distribution.

A useful measure of the collective layer dynamics is provided by the sum of
the maximum amplitudes over the layers, which hardly changes over time
(Figs.~\ref{fig:J15}c,d and \ref{fig:J25}c,d).

\subsection*{Analytical description of the layer distribution}

The connection between tunnelling and breather dynamics can be made explicit
analytically. When the radial soliton profile $\tilde{C}(K_{Z},R,\varphi,t)$
is independent of $K_{Z}$ (see above), the layer-dependent amplitude can be
factorized as
\begin{equation}
C(nd,R,\varphi,t)=\sqrt{\frac{2}{\tilde{N}+1}}\exp(-i\bar{W}_{0}t)\tilde
{C}(K_{Z},R,\varphi,t)A(J,n,t) \label{eq:C(nd,R,t)2}%
\end{equation}
using Eqs.(\ref{eq.:C_(K_Z,R,fi,t)}) and (\ref{eq:C(nd,R,t)}), where
\begin{equation}
A(J,n,t)=\sum_{K_{Z}}\exp[-i2Jt\cos\left(  \frac{\pi K_{Z}}{\tilde{N}%
+1}\right)  ]\sin\left(  \frac{\pi nK_{Z}}{\tilde{N}+1}\right)
\label{eq:A(J,n,t)}%
\end{equation}
This expression isolates the effect of interlayer tunnelling in the function
$A(J,n,t)$. The corresponding time dependence arises entirely from the
tunnelling-induced phase factor in the sum over $K_{Z}$.

Bearing in mind that $\exp[-i2Jt\cos\left(  \frac{\pi K_{Z}}{\tilde{N}%
+1}\right)  ]=\cos\left[  2Jt\cos\left(  \frac{\pi K_{Z}}{\tilde{N}+1}\right)
\right]  -i\sin\left[  2Jt\cos\left(  \frac{\pi K_{Z}}{\tilde{N}+1}\right)
\right]  $, using the expansions%
\begin{align*}
\cos(\zeta\cos\tau)  &  =J_{0}(\zeta)+2\sum_{p=1}^{\infty}(-1)^{p}J_{2p}%
(\zeta)\cos\left(  2p\tau\right)  ,\\
\sin(\zeta\cos\tau)  &  =-2\sum_{p=1}^{\infty}(-1)^{p}J_{2p\mathbf{-}1}%
(\zeta)\cos[(2p-1)\tau]
\end{align*}
where $J_{m}(\zeta)$ are the Bessel functions of the first kind, and summing
with respect to the discrete wave vectors $K_{Z}$, we get for the quantity
$A(J,n,t)$:%

\begin{align}
&  A(J,n,t)=[J_{0}(2Jt)\frac{\sin(\frac{\tilde{N}}{\tilde{N}+1}\frac{n\pi}%
{2})}{\sin\frac{n\pi}{2(\tilde{N}+1)}}+\sum_{p=1}^{\infty}(-1)^{p}J_{2p}(2Jt)%
%TCIMACRO{\dsum \limits_{+,\mathbf{-}}}%
%BeginExpansion
{\displaystyle\sum\limits_{+,\mathbf{-}}}
%EndExpansion
\frac{\sin(\tilde{N}\frac{\pi(n\pm2p)}{2(\tilde{N}+1)})}{\sin\frac{\pi
(n\pm2p)}{2(\tilde{N}+1)}}]\times\left\{
\begin{array}
[c]{c}%
1\text{, }n\text{ is odd}\\
0\text{, }n\text{ is even}%
\end{array}
\right\} \nonumber\\
&  +i\sum_{p=1}^{\infty}(-1)^{p}J_{2p\mathbf{-}1}(2Jt)%
%TCIMACRO{\dsum \limits_{+,\mathbf{-}}}%
%BeginExpansion
{\displaystyle\sum\limits_{+,\mathbf{-}}}
%EndExpansion
\frac{\sin(\tilde{N}\frac{\pi(n\pm(2p-1))}{2(\tilde{N}+1)})}{\sin\frac
{\pi(n\pm(2p-1))}{2(\tilde{N}+1)}}\times\left\{
\begin{array}
[c]{c}%
0\text{, }n\text{ is odd}\\
1\text{, }n\text{ is even}%
\end{array}
\right\}  \label{eq:A(J,n,t)2}%
\end{align}
Its real and imaginary parts have different parity with respect to the layer
index $n$: the real part vanishes for even $n$, whereas the imaginary part
vanishes for odd $n$.

\begin{figure}
%[ptbptbptbptbptbptbptbptbptbptbptb]
\begin{center}
\includegraphics[scale=0.784]{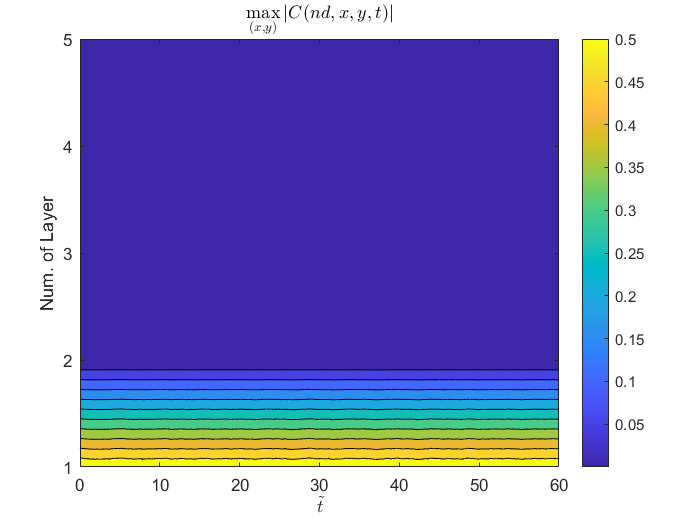}%
\caption{A maximum value of $C(nd,R,\varphi,t)$ as a function of time for
$J=0$ and $\tilde{N}=5$.}%
\label{fig:J=0}%
\end{center}
\end{figure}

The limiting case $J=0$ provides a particularly transparent test of this
interpretation. When interlayer tunnelling is absent, the phase factor becomes
independent of time, and
\begin{equation}
A(0,n,t)=\frac{\sin(\frac{\tilde{N}}{\tilde{N}+1}\frac{n\pi}{2})}{\sin
\frac{n\pi}{2(\tilde{N}+1)}}\times\left\{
\begin{array}
[c]{c}%
1\text{, }n\text{ is odd}\\
0\text{, }n\text{ is even}%
\end{array}
\right\}  , \label{eq:A(0,n,t)2}%
\end{equation}
Consequently, the layer distribution is stationary and the soliton does not
exhibit breather dynamics. The numerical result for $J=0$ shown in
Fig.~\ref{fig:J=0} confirms this behavior.

These results establish a direct correspondence between the presence of
interlayer tunnelling and the temporal modulation of the soliton in the layer
direction. Within the present model, the breather dynamics therefore provide a
potential signature of finite interlayer tunnelling.

\section*{DISCUSSION}

We have extended the soliton description of high-temperature SF in quasi-2D
perovskites to periodic SL structures. The resulting 3D nonlocal NLS equation
preserves the stable fundamental-soliton solution associated with the
exciton--LO phonon interaction, while the SL introduces a new degree of
freedom through the discrete out-of-plane momentum $K_{Z}$. The key new
feature is the miniband dispersion generated by interlayer tunnelling.

Our results show that the fundamental soliton retains its in-plane spatial
profile but acquires a $K_{Z}$-dependent phase through the superlattice
dispersion. Transforming this state to the layer representation converts the
phase modulation into a coherent temporal redistribution of the exciton
amplitude among the layers. The resulting state exhibits breather dynamics:
the soliton remains a stable nonlinear excitation, while its distribution
along the SL oscillates in time.

The physical origin of this behavior is therefore distinct from modulational
instability. The stability of the underlying fundamental soliton is inherited
from the nonlocal nonlinear interaction responsible for the stable excitonic
state, whereas the oscillatory dynamics along the SL arise from coherent
interlayer tunnelling and miniband formation. Increasing the number of layers
introduces additional miniband components and consequently produces a richer
spectrum of oscillation frequencies and beating dynamics.

The $J=0$ limit provides a direct way to distinguish these two effects. In the
absence of interlayer tunnelling, the miniband dispersion vanishes, the
$K_{Z}$-dependent phase modulation is absent, and the layer distribution
becomes time independent. The fundamental soliton therefore remains stationary
in the layer representation rather than forming a breather. Within the
assumptions of the present model, the emergence of breather dynamics thus
requires finite interlayer tunnelling.

These findings suggest an experimentally testable connection between the
microscopic coupling of neighboring perovskite layers and the collective
dynamics of the SF state. In particular, observation of coherent temporal
modulation of the emission or excitonic population associated with the layer
degree of freedom could provide evidence for finite interlayer tunnelling. The
characteristic frequencies and their evolution with the number of layers and
tunnelling strength further provide potential observables for probing the
miniband structure.

More broadly, the results demonstrate that the soliton mechanism proposed for
room-temperature SF in quasi-2D perovskites is not restricted to isolated
layers. SL engineering introduces a controllable out-of-plane degree of
freedom that transforms the stationary fundamental soliton into a dynamically
evolving breather through interlayer tunnelling. This provides a link between
exciton--phonon-induced nonlinear localization, miniband formation and
collective coherent dynamics in perovskite SLs.

\section*{METHODS}

We aim to extend a rigorous approach, based on fundamental principles, for
describing vibration-assisted single-exciton wavefunction evolution in
quasi-2D perovskites \cite{Fainberg_Osipov2024JCP} to the SL architecture. We
derive a set of equations for the evolution of the vibration-assisted exciton
wave-function, using the Hartree approximation. This derivation is similar to
those presented in Refs.\cite{Osipov_Fainberg23PRB,Fainberg_Osipov2024JCP}. We
use the coherent states $|\sigma\rangle$ \cite{Glauber63,BBGK1971} as the
basis for the phonon states
\cite{Osipov_Fainberg23PRB,Fainberg_Osipov2024JCP,s5h7-rpmk}. Each coherent
state $|\sigma\rangle$ is parametrized by a multidimensional complex-valued
vector $\sigma$ that encodes the coherent state center, i.e. the classical
coordinate $x$ and classical momentum $p$, namely $\sigma=x+ip$. The vibration
operators act on these basis vectors as follows: $b|\sigma\rangle
=\sigma|\sigma\rangle,\quad\langle\sigma|b^{\dag}=\langle\sigma|\sigma^{\ast}%
$. We follow the standard scheme used in the multiconfiguration Hartree
approach~\cite{MMC1992, WG2019, Osipov_Fainberg23PRB,Fainberg_Osipov2024JCP}.
The advantage of this approach is that the resulting dynamic equations are
much simpler for analysis than the original Schr\"{o}dinger equation; the
equations are similar to those of the mean-field Hartree theory. To describe
the wavefunction time evolution $|\Psi(t)\rangle$, we employ the
time-dependent basis vectors Ansatz. It assumes the basis vectors
$|\mathbf{\sigma}(t),\mathbf{q},K_{Z}\rangle$ depend on time in addition to
the time-dependent expansion coefficients $C_{0}(\mathbf{\sigma}%
,\mathbf{q},K_{Z}|t)$.

Thus, our working basis consists of the direct products of the exciton and
vibrational states%
\begin{equation}
|\mathbf{\sigma};\mathbf{q},K_{Z}\rangle=|\mathbf{\sigma}\rangle
|\mathbf{q},K_{Z}\rangle,\quad|\mathbf{\sigma}\rangle=\bigotimes_{\mathbf{q}%
}|\sigma_{\mathbf{q}}\rangle. \label{ketsigmaa}%
\end{equation}
With this, the wave function is expanded as follows:%

\begin{equation}
|\Psi(t)\rangle=\sum_{\mathbf{q}K_{Z}}[G(\mathbf{q},K_{Z}|t)|0\rangle
+\sum_{\mathbf{\sigma}}C_{0}(\mathbf{\sigma},\mathbf{q},K_{Z}%
|t)|\mathbf{\sigma}(t),\mathbf{q},K_{Z}\rangle] \label{wavefunction1}%
\end{equation}
To proceed, we use the Dirac--Frenkel variational principle, which is well
suited to describing quantum dynamics in systems with a large number of
vibrational degrees of freedom~\cite{SC2004, AR2017, Miller2002, WG2019}.
Variation of the wavefunction allows us to separate the time evolution of the
vibrational subsystem from the quantum dynamics of the exciton wavefunction.
The derivation of the resulting equations of motion follows a procedure
similar to that employed in our recent studies
\cite{Osipov_Fainberg23PRB,Fainberg_Osipov2024JCP}; however, the full
derivation is rather cumbersome. We therefore present a simplified derivation
based on the Heisenberg equations of motion, followed by the replacement of
the phonon operators with the corresponding coherent states, as in
Ref.\cite{s5h7-rpmk}. The Heisenberg equations of motion for the phonon
$b_{\mathbf{q}}$ and exciton $B_{\mathbf{q}}$ operators are
\begin{equation}
\dot{b}_{\mathbf{q}}=\frac{i}{\hbar}\left[  \hat{H}_{0},b_{\mathbf{q}}\right]
=-i\omega_{l}b_{\mathbf{q}}+D_{l}^{pol}(\mathbf{-q};00)\sum_{\mathbf{k}%
}B_{\mathbf{k}}^{\dag}B_{\mathbf{k+q}}, \label{eq:db_q/dt}%
\end{equation}

\begin{equation}
\dot{B}_{\mathbf{q}}=\frac{i}{\hbar}\left[  \hat{H}_{0},B_{\mathbf{q}}\right]
=-\frac{i}{\hbar}E_{exc}(\mathbf{k},K_{z})B_{\mathbf{q}}+\sum_{\mathbf{k}%
}D_{l}^{pol\ast}(\mathbf{k};00)(b_{\mathbf{k}}^{\dag}-b_{\mathbf{-k}%
})B_{\mathbf{q+k}} \label{eq:dB_nk/dt}%
\end{equation}
We average these equations using the single-exciton wavefunction,
Eq.(\ref{wavefunction1}), together with the normalization condition,
Eq.(\ref{eq:norm}). Replacing $b_{\mathbf{q}}$ with $\sigma_{\mathbf{k}}$
\cite{s5h7-rpmk} and, in the limit in which the ground-state population
remains close to unity, approximating the excitonic polarization $\langle
B_{\mathbf{q}}\rangle$ by the coefficient $C_{0}(\mathbf{q},K_{Z}|t)$ of the
single-exciton wavefunction, we obtain the equation of motion for
$\sigma_{\mathbf{q}}$
\begin{equation}
\dot{\sigma}_{\mathbf{q}}=-(i\omega_{l}+\gamma)\sigma_{\mathbf{q}}%
+D_{l}^{pol\ast}(\mathbf{q};00)\sum_{\mathbf{k}K_{Z}}C_{0}^{\ast}%
(\mathbf{k},K_{Z}|t)C_{0}(\mathbf{k+q},K_{Z}|t) \label{eq:dsigma_ks/dt}%
\end{equation}
In Eq.(\ref{eq:dsigma_ks/dt}), we also introduce a small decay $\gamma$ rate
of the mode $\omega_{l}$. This procedure transforms Eq.(\ref{eq:dB_nk/dt})
into a nonlinear equation of motion in momentum space
\begin{equation}
\dot{C}_{0}(\mathbf{q},K_{Z}|t)=-\frac{i}{\hbar}E_{exc}(\mathbf{k},K_{z}%
)C_{0}(\mathbf{q},K_{Z}|t)+\sum_{\mathbf{k}}\tilde{\alpha}_{l}(\mathbf{k}%
;00)C_{0}(\mathbf{q}+\mathbf{k},K_{Z}|t), \label{eq.:dC_0(q,nd,t)/dt}%
\end{equation}
in which the exciton--LO phonon interaction enters through a mean-field
Hartree term $\tilde{\alpha}_{l}\mathbf{(k};00)\mathbf{=}D_{l}^{pol\ast
}\mathbf{(k};00)(\sigma_{\mathbf{k}}^{\ast}-\sigma_{\mathbf{-k}})$. Bearing in
mind that mode $\omega_{l}$ is long-lived LO phonon mode and may be associated
with lead--halide--lead rocking vibrations \cite{Gundogdu2025Nature}, similar
to Eq.(8) of Ref.\cite{s5h7-rpmk}, parameter $\tilde{\alpha}_{l}%
\mathbf{(k};00)$ can be approximated by%

\[
\tilde{\alpha}_{l}(\mathbf{k};00)\approx i2\omega_{0}(k)\sum_{\mathbf{q}K_{Z}%
}C_{0}^{\ast}(\mathbf{q},K_{Z}|t)C_{0}(\mathbf{q}-\mathbf{k},K_{Z}|t)
\]
where $\omega_{0}(k)=\frac{\omega_{l}}{\omega_{l}^{2}+\gamma^{2}}|D_{l}%
^{pol}(\mathbf{k};00)|^{2}$. Then we get from Eq.(\ref{eq.:dC_0(q,nd,t)/dt})%
\begin{equation}
\dot{C}_{0}(\mathbf{q},K_{Z}|t)=-\frac{i}{\hbar}E_{exc}(\mathbf{k},K_{z}%
)C_{0}(\mathbf{q},K_{Z}|t)+\sum_{\mathbf{k}}i2\omega_{0}(k)\sum_{\mathbf{q}%
^{\prime}K_{Z}^{\prime}}C_{0}^{\ast}(\mathbf{q}^{\prime},K_{Z}^{\prime
}|t)C_{0}(\mathbf{q}^{\prime}+\mathbf{k},K_{Z}^{\prime}|t)C_{0}(\mathbf{q}%
+\mathbf{k},K_{Z}|t) \label{eq.:dC_0(q,K_Z,t)/dt}%
\end{equation}
Eq.(\ref{eq.:dC_0(q,K_Z,t)/dt}) represents a nonlinear equation for
coefficients $C_{0}(\mathbf{q},K_{Z}|t)$ in the momentum space.

\section*{DATA AVAILABILITY}

The data that support the findings of this study are available from the
corresponding author upon reasonable request; they are not publicly available
because they are owned by a third party and the terms of use prevent public distribution.

\section*{CODE AVAILABILITY}

The code used for the numerical calculations is available from the
corresponding author upon reasonable request.

\section*{REFERENCES}

%\bibliographystyle{IEEEtran}
%\bibliography{ADV2,ADV3,ADVCHEM}

\begin{thebibliography}{99}                                                                                               %
\providecommand{\url}[1]{#1}

\providecommand{\newblock}{\relax} \providecommand{\bibinfo}[2]{#2}
\providecommand\BIBentrySTDinterwordspacing{\spaceskip=0pt\relax}
\providecommand\BIBentryALTinterwordstretchfactor{4}
\providecommand\BIBentryALTinterwordspacing{\spaceskip=\fontdimen2\font plus
\BIBentryALTinterwordstretchfactor\fontdimen3\font minus
\fontdimen4\font\relax}
\providecommand\BIBforeignlanguage[2]{{\expandafter\ifx\csname l@#1\endcsname\relax
\typeout{** WARNING: IEEEtran.bst: No hyphenation pattern has been}\typeout{** loaded for the language `#1'. Using the pattern for}\typeout{** the default language instead.}\else
\language=\csname l@#1\endcsname
\fi
#2}}

\bibitem {Gundogdu2022Nature_Phot}M.~Biliroglu, G.~Findik, J.~Mendes,
D.~Seyitliyev, L.~Lei, Q.~Dong, Y.~Mehta, V.~V. Temnov, F.~So, and
K.~Gundogdu, ``Room-temperature superfluorescence in hybrid perovskites and
its origins,'' \emph{Nature Photonics}, vol.~16, pp. 324--329, 2022.

\bibitem {Sum_Mhaisalkar_Bruno2026AdvMat}Y.~Tang, H.~T. Ching, G.~Ong,
C.~Kulshreshtha, Z.~Xing, H.~A. Dewi, L.~R.~W. White, K.~J. How, L.~C. Kwek,
T.~C. Sum, S.~G. Mhaisalkar, and A.~Bruno, ``Low-threshold superfluorescence
and phase dynamics in quasi-2d metal halide perovskite thin films,''
\emph{Adv. Mater.}, vol.~38, p. e18842, 2026.

\bibitem {Gundogdu2025Nature}M.~Biliroglu, M.~Turel, A.~Ghita, M.~Kotyrov,
X.~Qin, D.~Seyitliyev, N.~Phonthiptokun, M.~Abdelsamei, J.~Chai, R.~Su,
U.~Herath, A.~K. Swan, V.~V. Temnov, V.~Blum, F.~So, and K.~Gundogdu,
``Unconventional solitonic high-temperature superfluorescence from
perovskites,'' \emph{Nature}, vol. 642, pp. 71--78, 2025.

\bibitem {s5h7-rpmk}A.~A. Gladkij, N.~A. Veretenov, N.~N. Rosanov, B.~A.
Malomed, V.~A. Osipov, and B.~D. Fainberg, ``Stability of the quantum coherent
superradiant states in relation to exciton-phonon interactions and the
fundamental soliton in hybrid perovskites,'' \emph{Phys. Rev. B}, vol. 113, p.
195439, 2026, [arXiv:2511.03600 (2025)].

\bibitem {Zeng2026Materials_Futures}W.~Zeng, J.~Xu, and X.~Chen, ``A new
perspective in understanding the room-temperature superfluorescence from
perovskites: collectively coherent excitonic polaron state,'' \emph{Mater.
Futures}, vol.~5, p. 037501, 2026.

\bibitem {Fainberg_Osipov2024JCP}B.~D. Fainberg and V.~A. Osipov, ``Theory of
high-temperature superfluorescence in hybrid perovskite thin films,'' \emph{J.
Chem. Phys.}, vol. 161, no.~11, p. 114705, 2024.

\bibitem {Miyata17}K.~Miyata, D.~Meggiolaro, M.~T. Trinh, P.~P. Joshi,
E.~Mosconi, S.~C. Jones, F.~D. Angelis, and X.-Y. Zhu, ``Large polarons in
lead halide perovskites,'' \emph{Sci. Adv.}, vol.~3, p. e1701217, 2017.

\bibitem {Osipov_Fainberg23PRB}V.~A. Osipov and B.~Fainberg, ``Hartree method
for molecular polaritons,'' \emph{Phys. Rev. B}, vol. 107, p. 075404, 2023.

\bibitem {Krolikowski2000}W.~Krolikowski and O.~Bang, ``Solitons in nonlocal
nonlinear media: Exact solutions,'' \emph{Phys. Rev. E}, vol.~63, p. 016610, 2000.

\bibitem {Malomed2022}B.~A. Malomed, ``Two-dimensional solitons in nonlocal
media: A brief review,'' \emph{Symmetry}, vol.~14, p. 1565, 2022.

\bibitem {Krolikowski2001}W.~Krolikowski, O.~Bang, J.~J. Rasmussen, and
J.~Wyller, ``Modulational instability in nonlocal nonlinear kerr media,''
\emph{Phys. Rev. E}, vol.~64, p. 016612, 2001.

\bibitem {Krolikowsk2004}W.~Krolikowski, ``Modulational instability, solitons
and beam propagation in spatially nonlocal nonlinear media,'' \emph{J. Opt. B:
Quantum Semiclass. Opt.}, vol.~6, pp. S288--S294, 2004.

\bibitem {Raino2018Nature}G.~Raino, M.~A. Becker, M.~I. Bodnarchuk, R.~F.
Mahrt, M.~V. Kovalenko, and T.~Stoferle, ``Superfluorescence from lead halide
perovskite quantum dot superlattices,'' \emph{Nature}, vol. 563, pp. 671--675, 2018.

\bibitem {Cherniukh2021Nature}I.~Cherniukh, G.~Raino, T.~Stoferle, M.~Burian,
A.~Travesset, D.~Naumenko, H.~Amenitsch, R.~Erni, R.~F. Mahrt, M.~I.
Bodnarchuk, and M.~V. Kovalenko, ``Perovskite-type superlattices from lead
halide perovskite nanocubes,'' \emph{Nature}, vol. 593, pp. 535--542, 2021.

\bibitem {Wildenborg2025ACSPhotonics}A.~J. Wildenborg, R.~J. Munter,
F.~Freire-Fernandez, E.~T.~F. Freitas, and J.~Y. Suh, ``Superlattice-induced
superfluorescence in quasi-2d metal halide perovskites,'' \emph{ACS
Photonics}, vol.~12, pp. 3476--3483, 2025.

\bibitem {Mandal2026Nat_Commun}S.~R. Koshkaki, A.~Manjalingal, L.~Blackham,
and A.~Mandal, ``Exciton-polariton dynamics in multilayered materials,''
\emph{Nat. Commun.}, vol.~17, p. 6156, 2026.

\bibitem {Fra98}E.~Fradkin, \emph{Field theories of condensed matter
systems}.\hskip 1em plus 0.5em minus 0.4em\relax New York: Addison-Wesley, 1991.

\bibitem {Che63}D.~B. Chesnut and A.~Suna, ``Fermion behavior of
one-dimensional excitons,'' \emph{J. Chem. Phys.}, vol.~39, no.~1, pp.
146--149, 1963.

\bibitem {Spano91}F.~C. Spano, ``Fermion excited states in one-dimensional
molecular aggregates with site disorder: nonlinear optical response,''
\emph{Phys. Rev. Lett.}, vol.~24, no.~24, pp. 3424--3427, 1991.

\bibitem {Muk95}S.~Mukamel, \emph{Principles of Nonlinear Optical
Spectroscopy}.\hskip 1em plus 0.5em minus 0.4em\relax New York: Oxford
University Press, 1995.

\bibitem {Fainberg2016CP}B.~N. Levinsky, B.~D. Fainberg, L.~A. Nesterov, and
N.~N. Rosanov, ``Two-exciton excited states of j-aggregates in the presence of
exciton-exciton annihilation,'' \emph{Chem. Phys.}, vol. 473, pp. 1--10, 2016.

\bibitem {Glauber63}R.~J. Glauber, ``Coherent and incoherent states of the
radiation field,'' \emph{Phys. Rev}, vol. 131, no.~2, pp. 2766--2788, 1963.

\bibitem {BBGK1971}J.-M. Sixdeniers and K.~A. Penson, ``On the completeness of
coherent states generated by binomial distribution,'' \emph{Journal of Physics
A: Mathematical and General}, vol.~33, no.~14, pp. 2907--2916, 2000.

\bibitem {MMC1992}U.~Manthe, H.~Meyer, and L.~S. Cederbaum, \textquotedblleft
Wave packet dynamics within the multiconfiguration Hartree framework: General
aspects and application to NOCl,\textquotedblright\ \emph{The Journal of
Chemical Physics}, vol.~97, no.~5, pp. 3199--3213, 1992.

\bibitem {WG2019}M.~Werther and F.~Gro\ss {}mann, ``Apoptosis of moving
nonorthogonal basis functions in many-particle quantum dynamics,'' \emph{Phys.
Rev. B}, vol. 101, p. 174315, 2020.

\bibitem {SC2004}D.~V. Shalashilin and M.~S. Child, \textquotedblleft The
phase space ccs approach to quantum and semiclassical molecular dynamics for
high-dimensional systems,\textquotedblright\ \emph{Chemical Physics}, vol.
304, no.~1, pp. 103--120, 2004.

\bibitem {AR2017}E.~Artacho and D.~D. O'Regan, \textquotedblleft Quantum
mechanics in an evolving hilbert space,\textquotedblright\ \emph{Phys. Rev.
B}, vol.~95, p. 115155, 2017.

\bibitem {Miller2002}W.~H. Miller, ``On the relation between the semiclassical
initial value representation and an exact quantum expansion in time-dependent
coherent states,'' \emph{The Journal of Physical Chemistry B}, vol. 106,
no.~33, pp. 8132--8135, 2002.
\end{thebibliography}

\section*{ACKNOWLEDGEMENTS}

B.F. acknowledges support by the European Cooperation in Science and
Technology (COST Action No. CA24109-QOpen (Many-body Open Quantum Systems)).
A.A.G. and N.N.R. were supported by Ioffe Institute State Assignment, topic 0040-2019-0017.

\section*{AUTHOR CONTRIBUTIONS}

A.A.G. was responsible for formal analysis, software, validation, and
visualization. N.N.R. was responsible for conceptualization, formal analysis,
methodology, software, supervision, validation, and visualization. B.D.F. was
responsible for conceptualization, formal analysis, methodology, project
administration, validation, visualization, writing (original draft), review,
and editing.

\section*{COMPETING INTERESTS}

The authors declare no competing interests.%

%TCIMACRO{\FRAME{ftbpFU}{6.6331in}{3.5155in}{0pt}{\Qcb{On the left, radially
%symmetric superlattice profiles of $\tilde{C}(K_{Z},R,t)$ for $\tilde{N}=15$.
%On the right is a plot showing the superposition of all $K_{Z}$
%distributions.}}{\Qlb{fig:C(K_Z,R,t)}}{fig1.png}%
%width 6.6331in;  height 3.5155in;  depth 0pt;  original-width 14.6767in;
%original-height 7.7496in;  cropleft "0";  croptop "1";  cropright "1";
%cropbottom "0";  filename 'Fig1.png';file-properties "XNPEU";}}}%
%BeginExpansion
\begin{figure}
%[ptb]
\begin{center}
\includegraphics[scale=0.45]{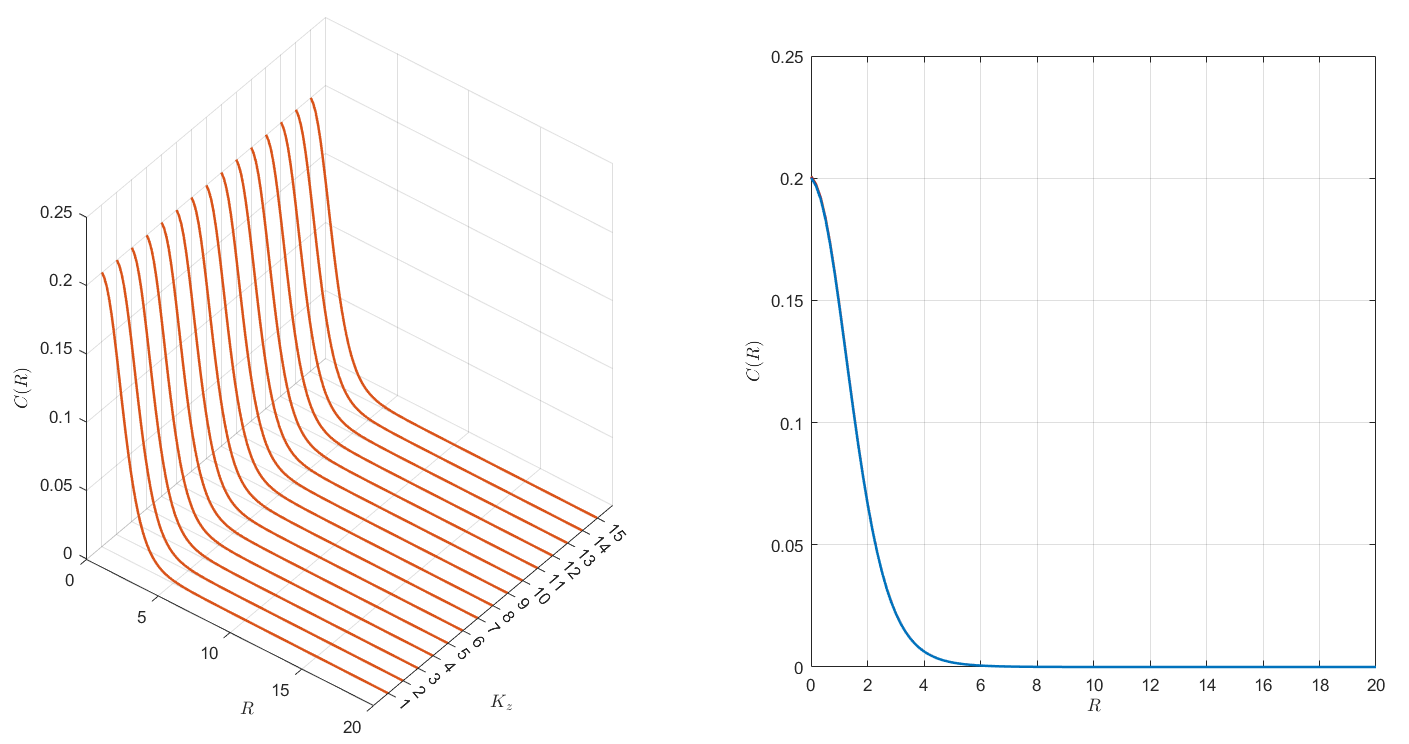}
\caption{On the left, radially symmetric SL profiles of $\tilde{C}(K_{Z},R,t)$ 
for $\tilde{N}=15$; on the right is a plot showing the
superposition of all $K_{Z}$ distributions.}%
\label{fig:C(K_Z,R,t)}%
\end{center}
\end{figure}

\begin{figure}
%[ptbptb]
\begin{center}
\includegraphics[scale=0.488]{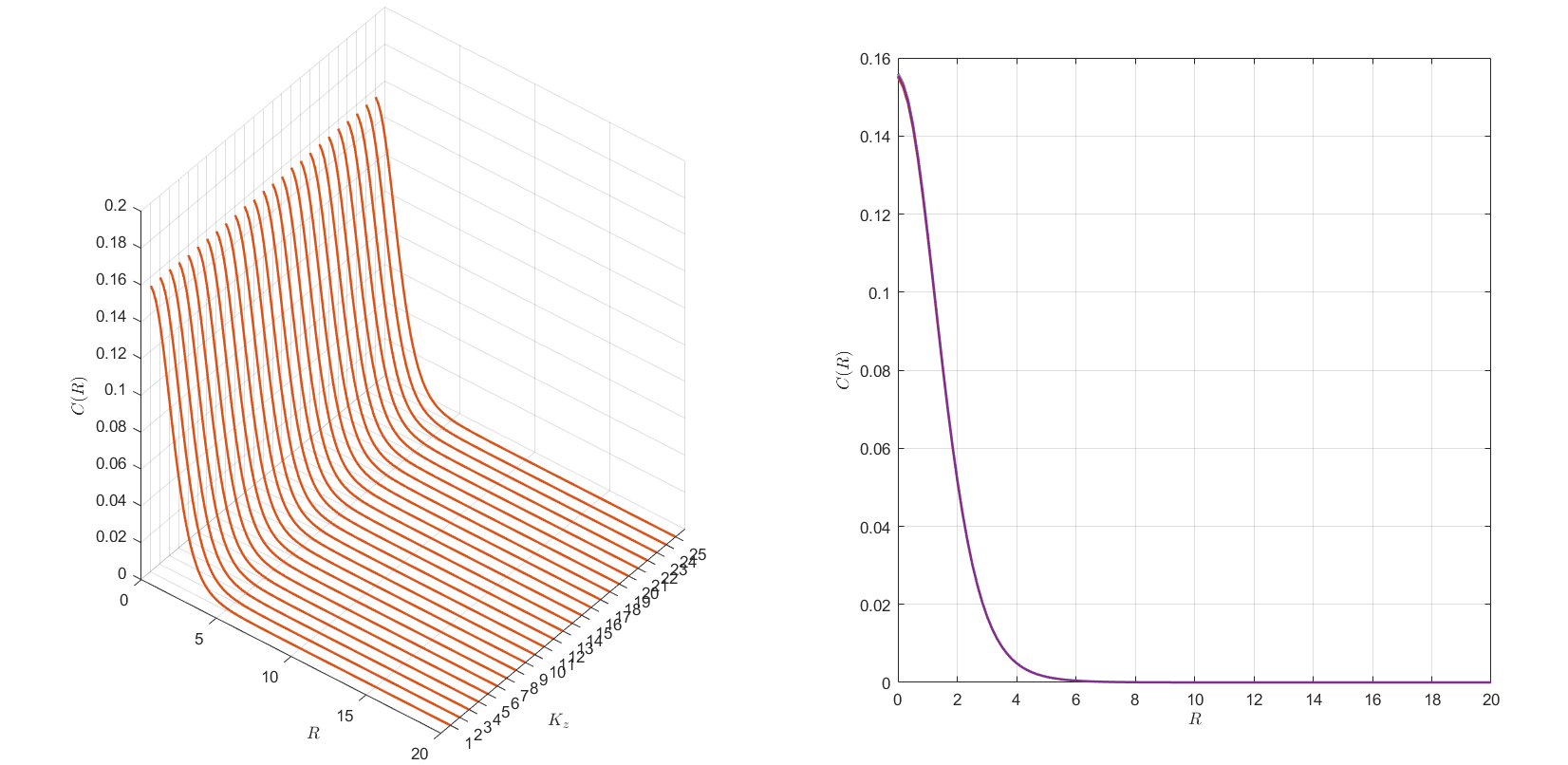}%
\caption{On the left, radially symmetric superlattice profiles of 
$\tilde{C}(K_{Z},R,t)$ for $\tilde{N}=25$. On the right is a plot showing the
superposition of all $K_{Z}$ distributions.}%
\label{fig:C(K_Z,R,t)25}%
\end{center}
\end{figure}
%EndExpansion%
%TCIMACRO{\FRAME{ftbpFU}{3.9764in}{1.9501in}{0pt}{\Qcb{Frenkel exciton model of
%SL with period $d$.}}{\Qlb{fig:Frenkel}}{frenkel.png}%
%width 3.9764in;  height 1.9501in;  depth 0pt;  original-width 12.6877in;
%original-height 6.1774in;  cropleft "0";  croptop "1";  cropright "1";
%cropbottom "0";  filename 'Frenkel.png';file-properties "XNPEU";}}}%
%BeginExpansion
%EndExpansion%
%TCIMACRO{\FRAME{ftbpFU}{5.3765in}{1.9493in}{0pt}{\Qcb{A maximum value of
%$C(nd,R,\varphi,t)$ as a function of time for $J=2000$ $cm^{\mathbf{-}1}$and
%$\tilde{N}=3$.}}{\Qlb{fig:C(nd,R,t)}}{fig2.png}%
%width 5.3765in;  height 1.9493in;  depth 0pt;  original-width 15.8537in;
%original-height 5.6982in;  cropleft "0";  croptop "1";  cropright "1";
%cropbottom "0";  filename 'Fig2.png';file-properties "XNPEU";}}}%
%BeginExpansion
\begin{figure}
%[ptbptbptbptb]
\begin{center}
\includegraphics[scale=0.342]{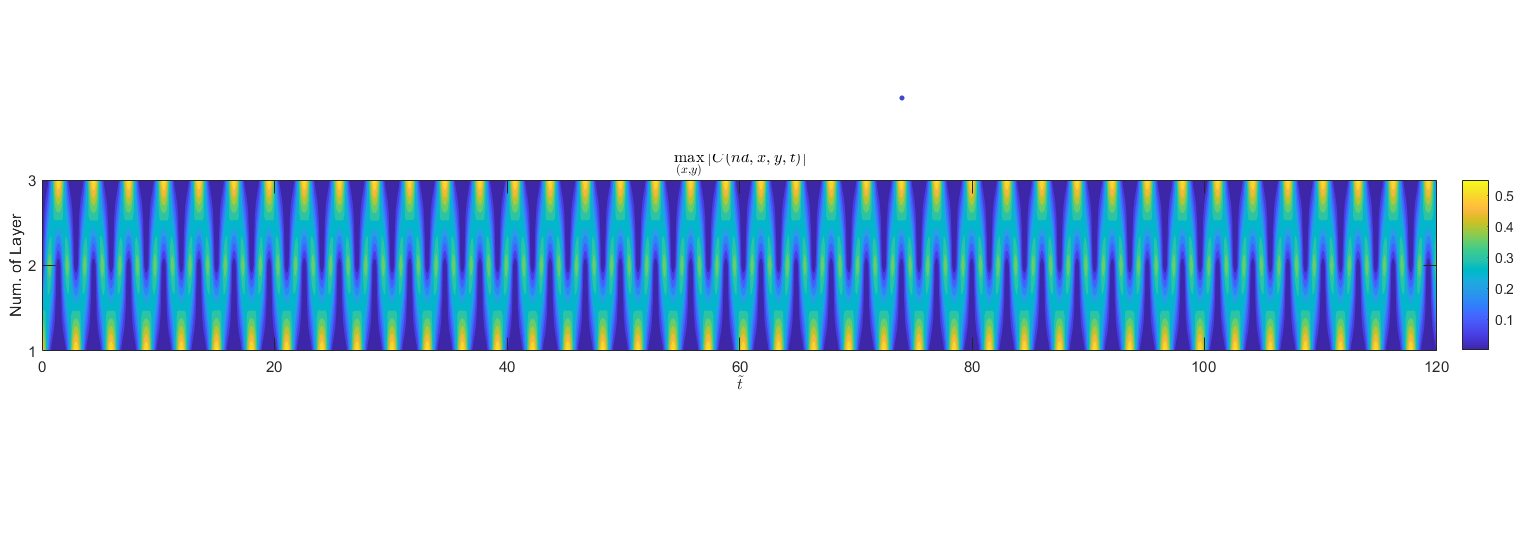}%
\caption{A maximum value of $C(nd,R,\varphi,t)$ as a function of time for
$J=2000$ $cm^{\mathbf{-}1}$and $\tilde{N}=3$.}%
\label{fig:C(nd,R,t)}%
\end{center}
\end{figure}
%EndExpansion

%EndExpansion%
%TCIMACRO{\FRAME{ftbpFU}{6.0061in}{0.9729in}{0pt}{\Qcb{A maximum value of
%$C(nd,R,\varphi,t)$ as a function of time for $J=2000$ $cm^{\mathbf{-}1}$and
%$\tilde{N}=5$.}}{\Qlb{fig:5layers}}{fig5layers.png}%
%width 6.0061in;  height 0.9729in;  depth 0pt;  original-width 15.6773in;
%original-height 2.4794in;  cropleft "0";  croptop "1";  cropright "1";
%cropbottom "0";  filename 'Fig5layers.png';file-properties "XNPEU";}}}%
%BeginExpansion
%EndExpansion%
%TCIMACRO{\FRAME{ftbpFU}{6.0079in}{1.1398in}{0pt}{\Qcb{A maximum value of
%$C(nd,R,\varphi,t)$ as a function of time for $J=2000$ $cm^{\mathbf{-}1}$and
%$\tilde{N}=15$.}}{\Qlb{fig:15layers}}{fig15layers.png}%
%width 6.0079in;  height 1.1398in;  depth 0pt;  original-width 15.5623in;
%original-height 2.8963in;  cropleft "0";  croptop "1";  cropright "1";
%cropbottom "0";  filename 'Fig15layers.png';file-properties "XNPEU";}}}%
%BeginExpansion
%EndExpansion%
%TCIMACRO{\FRAME{ftbpFU}{6.4065in}{4.1364in}{0pt}{\Qcb{Average coordinate
%$\langle Z(t)\rangle$ as a function of time (a) and the spectrum of $\langle
%Z(t)\rangle$ (b) for $J=2000$ $cm^{\mathbf{-}1}$and $\tilde{N}=5$.}%
%}{\Qlb{fig:Breather5}}{fig5layerszspectrum.png}%
%width 6.4065in;  height 4.1364in;  depth 0pt;  original-width 15.948in;
%original-height 10.2705in;  cropleft "0";  croptop "1";  cropright "1";
%cropbottom "0";  filename 'Fig5layersZspectrum.png';file-properties "XNPEU";}%
%}}%
%BeginExpansion
%EndExpansion%
%TCIMACRO{\FRAME{ftbpFU}{6.7213in}{4.3336in}{0pt}{\Qcb{Average coordinate
%$\langle Z(t)\rangle$ as a function of time (a) and the spectrum of $\langle
%Z(t)\rangle$ (b) for $J=2000$ $cm^{\mathbf{-}1}$and $\tilde{N}=15$.}%
%}{\Qlb{fig:Breather15}}{breather15.png}{\special{ language "Scientific Word";
%type "GRAPHIC";  maintain-aspect-ratio TRUE;  display "USEDEF";
%valid_file "F";  width 6.7213in;  height 4.3336in;  depth 0pt;
%original-width 19.1253in;  original-height 12.302in;  cropleft "0";
%filename 'Breather15.png';file-properties "XNPEU";}}}%
%BeginExpansion

%EndExpansion%
%TCIMACRO{\FRAME{ftbpFU}{3.3615in}{3.3615in}{0pt}{\Qcb{Energy dependence on
%coordinate $Z$ for $\tilde{N}=5$.}}{\Qlb{fig:E_n5}}{n5.png}%
%width 3.3615in;  height 3.3615in;  depth 0pt;  original-width 3.333in;
%original-height 3.333in;  cropleft "0";  croptop "1";  cropright "1";
%cropbottom "0";  filename 'N5.png';file-properties "XNPEU";}}}%
%BeginExpansion
%EndExpansion%
%TCIMACRO{\FRAME{ftbpFU}{7.005in}{4.3189in}{0pt}{\Qcb{A maximum value of
%$C(nd,R,\varphi,t)$ as a function of time for $J=1000$ $cm^{\mathbf{-}1}$ (a)
%and $J=2000$ $cm^{\mathbf{-}1}$ (b) when $\tilde{N}=15$. Dependence of the sum
%of the maximum values of $C(nd,R,\varphi,t)$ over the layers as a function of
%time for $J=1000$ $cm^{\mathbf{-}1}$ (c) and $J=2000$ $cm^{\mathbf{-}1}$
%(d).}}{\Qlb{fig:J15}}{figj15.png}{\special{ language "Scientific Word";
%type "GRAPHIC";  maintain-aspect-ratio TRUE;  display "USEDEF";
%valid_file "F";  width 7.005in;  height 4.3189in;  depth 0pt;
%original-width 12.6877in;  original-height 7.8023in;  cropleft "0";
%filename 'FigJ15.png';file-properties "XNPEU";}}}%
%BeginExpansion
%EndExpansion%
%TCIMACRO{\FRAME{ftbpFU}{5.4959in}{4.2367in}{0pt}{\Qcb{A maximum value of
%$C(nd,R,\varphi,t)$ as a function of time for $J=0$ and $\tilde{N}=5$.}%
%}{\Qlb{fig:J=0}}{figj0.png}{\special{ language "Scientific Word";
%type "GRAPHIC";  maintain-aspect-ratio TRUE;  display "USEDEF";
%valid_file "F";  width 5.4959in;  height 4.2367in;  depth 0pt;
%original-width 7.0102in;  original-height 5.3956in;  cropleft "0";
%filename 'FigJ0.png';file-properties "XNPEU";}}}%
%BeginExpansion
%EndExpansion

\end{document}